\documentclass[]{article}
\usepackage{graphicx}
\usepackage{color}
\usepackage{listings}
\usepackage{fullpage}
\usepackage{amsmath}
\usepackage[utf8x]{inputenc}
\usepackage{import}
\usepackage{setspace}
\usepackage{authblk}
\usepackage{hyperref}
\usepackage{float}
\definecolor{lightgray}{gray}{0.5}
\usepackage{setspace}
\usepackage{color}
\usepackage[font={small,sf}, singlelinecheck=false]{caption}

\newcommand{\rev}[1]{\textcolor{black}{#1}}
\newcommand{\revtwo}[1]{\textcolor{black}{#1}}

\begin{document}

	\title{Efficient Coding in the Economics of Human Brain Connectomics}
	
	\author[1]{Dale Zhou}
	\author[2,3]{Christopher W. Lynn}
	\author[4]{Zaixu Cui}
	\author[5]{Rastko Ciric}
	\author[6]{Graham L. Baum}
	\author[4,7]{Tyler M. Moore}
	\author[4,7]{David R. Roalf}
	\author[8]{John A. Detre}
	\author[4,7]{Ruben C. Gur}
	\author[4,7]{Raquel E. Gur}
	\author[4,7 $\dagger$]{Theodore D. Satterthwaite} 
	\author[4,8,9,10,11,12,13 $\dagger$]{Dani S. Bassett}
	
	\affil[1]{Department of Neuroscience, Perelman School of Medicine, University of Pennsylvania, Philadelphia, PA 19104, USA}
	\affil[2]{Initiative for the Theoretical Sciences, Graduate Center, City University of New York, New York, NY 10016, USA}
	\affil[3]{Joseph Henry Laboratories of Physics, Princeton University, Princeton, NJ 08544, USA}
	\affil[4]{Department of Psychiatry, Perelman School of Medicine, University of Pennsylvania, Philadelphia, PA 19104, USA}
	\affil[5]{Department of Bioengineering, Schools of Engineering and Medicine, Stanford University, Stanford, CA 94305 USA}
	\affil[6]{Department of Psychology and Center for Brain Science, Harvard University, Cambridge, MA 02138, USA}
	\affil[7]{Penn-Children's Hospital of Philadelphia Lifespan Brain Institute, Philadelphia, PA 19104, USA}
	\affil[8]{Department of Neurology, Perelman School of Medicine, University of Pennsylvania, Philadelphia, PA 19104, USA}
	\affil[9]{Department of Physics \& Astronomy, College of Arts and Sciences, University of Pennsylvania, Philadelphia, PA 19104, USA}
	\affil[10]{Department of Bioengineering, School of Engineering and Applied Sciences, University of Pennsylvania, Philadelphia, PA 19104, USA}
	\affil[11]{Department of Electrical \& Systems Engineering, School of Engineering and Applied Sciences, University of Pennsylvania, Philadelphia, PA 19104, USA}
	\affil[12]{Santa Fe Institute, Santa Fe, NM 87501, USA}
	\affil[13]{To whom correspondence should be addressed: dsb@seas.upenn.edu}
	\affil[$\dagger$]{Co-senior authors}
	
	\maketitle
	
	\vspace*{\fill}
	
	\newpage
	
	\doublespacing
	
    \begin{center}
        \section*{Abstract}
    \end{center}

	In systems neuroscience, most models posit that brain regions communicate information under constraints of efficiency. Yet, evidence for efficient communication in structural brain networks characterized by hierarchical organization and highly connected hubs remains sparse. The principle of efficient coding proposes that the brain transmits maximal information in a metabolically economical or compressed form to improve future behavior. To determine how structural connectivity supports efficient coding, we develop a theory specifying minimum rates of message \rev{transmission} between brain regions to achieve an expected fidelity, and we test five predictions from the theory based on random walk communication dynamics. In doing so, we introduce the metric of compression efficiency, which quantifies the trade-off between lossy compression and transmission fidelity in structural networks. In a large sample of youth (\textit{n} = 1,042; age 8-23 years), we analyze structural networks derived from diffusion weighted imaging and metabolic expenditure operationalized using cerebral blood flow. We show that structural networks strike compression efficiency trade-offs consistent with theoretical predictions. We find that compression efficiency prioritizes fidelity with development, heightens when metabolic resources and myelination guide communication, explains advantages of hierarchical organization, links higher input fidelity to disproportionate areal expansion, and shows that hubs integrate information by lossy compression. Lastly, compression efficiency is predictive of behavior---beyond the conventional network efficiency metric---for cognitive domains including executive function, memory, complex reasoning, and social cognition. Our findings elucidate how macroscale connectivity supports efficient coding, and serve to foreground communication processes that utilize random walk dynamics constrained by network connectivity. \\

	\textbf{Author Summary.} Macroscale communication between interconnected brain regions underpins most aspects of brain function and incurs substantial metabolic cost. Understanding efficient and behaviorally meaningful information transmission dependent on structural connectivity has remained challenging. We validate a model of communication dynamics atop the macroscale human structural connectome, finding that structural networks support dynamics that strike a balance between information transmission fidelity and lossy compression. Notably, this balance is predictive of behavior and explanatory of biology. In addition to challenging and reformulating the currently held view that communication occurs by routing dynamics along metabolically efficient direct anatomical pathways, our results suggest that connectome architecture and behavioral demands yield communication dynamics that accord to neurobiological and information theoretical principles of efficient coding and lossy compression. \\
	
	\textbf{Keywords.} network communication dynamics, metabolic resources, lossy compression, rate-distortion theory, integration, hierarchical organization, network hubs
	
	\newpage
	
	\section*{Introduction}
	
	The principle of compensation states that ``to spend on one side, nature is forced to economise on the other side'' \cite{west2003developmental}. In the economics of brain connectomics, natural selection optimizes network architecture for versatility, resilience, and efficiency under constraints of metabolism, materials, space, and time \cite{laughlin2001energy, west2003developmental, bullmore2012economy}. Brain networks---composed of nodes representing cortical regions and edges representing white matter tracts---strike evolutionary compromises between costs and adaptations \cite{laughlin2001energy, west2003developmental, buckner2013evolution, avena2015network, whitaker2016adolescence, reardon2018normative}. Further, network disruptions may contribute to the development of neuropsychiatric disorders \cite{di2014unraveling, crossley2014hubs, gollo2018fragility, kaczkurkin2018common}. \rev{Because the limits of computations are intertwined with the limits of communication between brain regions \cite{cover1999elements}, to understand how the brain efficiently balances resource constraints with pressures of information processing, one must begin with models of information transmission in brain networks.} \\

    \rev{Principles of neurotransmission established at the cellular level suggest that biophysical constraints on information processing may apply to the macroscopic levels of brain regions and networks \cite{10.2307/j.ctt17kk982, levy1996energy, laughlin2001energy}. The principle of efficient coding proposes that the brain transmits maximal information in a metabolically economical or \textit{compressed} form to improve future behavior \cite{chalk2018toward}. Efficient coding at the macroscale offers a parsimonious principle of compression characterizing the dimensionality of neural representations \cite{tang2019effective, shine2019human, Stringer2019HighdimensionalGO, mack2020ventromedial}, as well as a parsimonious principle of transmission characterizing a spectrum of network communication mechanisms \cite{bullmore2012economy, goni2013exploring, goni2014resting, avena2015network, avena2017path, mivsic2015cooperative, avena2018communication}. However, it remains incompletely understood how this principle generalizes from cells and sensory systems to the macroscale connectome \cite{10.2307/j.ctt17kk982, chalk2018toward}.} \\
    
    An unexplored link between the efficient coding of compressed transmissions and macroscale brain network communication dynamics is rate-distortion theory, a major branch of information theory that establishes the mathematical foundations of \textit{lossy data compression} for any communication channel \cite{shannon1959coding}. Rate-distortion theory formalizes the link between compression and communication by determining the minimum amount of information that a source should transmit (the rate) for a target to approximately receive the input signal without exceeding an expected amount of noise (the distortion) \cite{shannon1959coding}. Lossy data compression is reducing the amount of information transmitted (rate), accepting some loss of data fidelity (distortion). Using a rate-distortion model, we sought to explain how the macroscale connectome supports efficient coding from minimal assumptions. \\
    
    \rev{We modeled information transmission as the passing of stochastic messages in parallel along the wiring of the human connectome (\textbf{Figure 1A}). More precisely, we modeled information transmission using a repetition code \cite{barlow1961possible, cover1999elements}. A repetition code uses redundancy---here by sending multiple copies of a message---to overcome errors in communication arising from the stochasticity of neural processes \cite{10.2307/j.ctt17kk982, cover1999elements}. A natural trade-off emerges between the redundancy and efficiency of a message: while redundant messages are more robust to errors in transmission, they also incur greater cost \cite{barlow1961possible}. Thus, given an allowed error rate (or, equivalently, an expected fidelity), maximizing the efficiency of information transmission requires minimizing the redundancy of messages \cite{cover1999elements}. By quantifying the minimal number of repeated messages needed to achieve a given fidelity, questions of connectome computation and communication can be formulated as a tractable mathematical problem using stochastic processes and redundancy reduction \cite{10.2307/j.ctt17kk982, avena2018communication, sims2018efficient}.}\\ 
    
	\begin{figure}[htbp!]
	\centering
	\includegraphics[width=1\columnwidth]{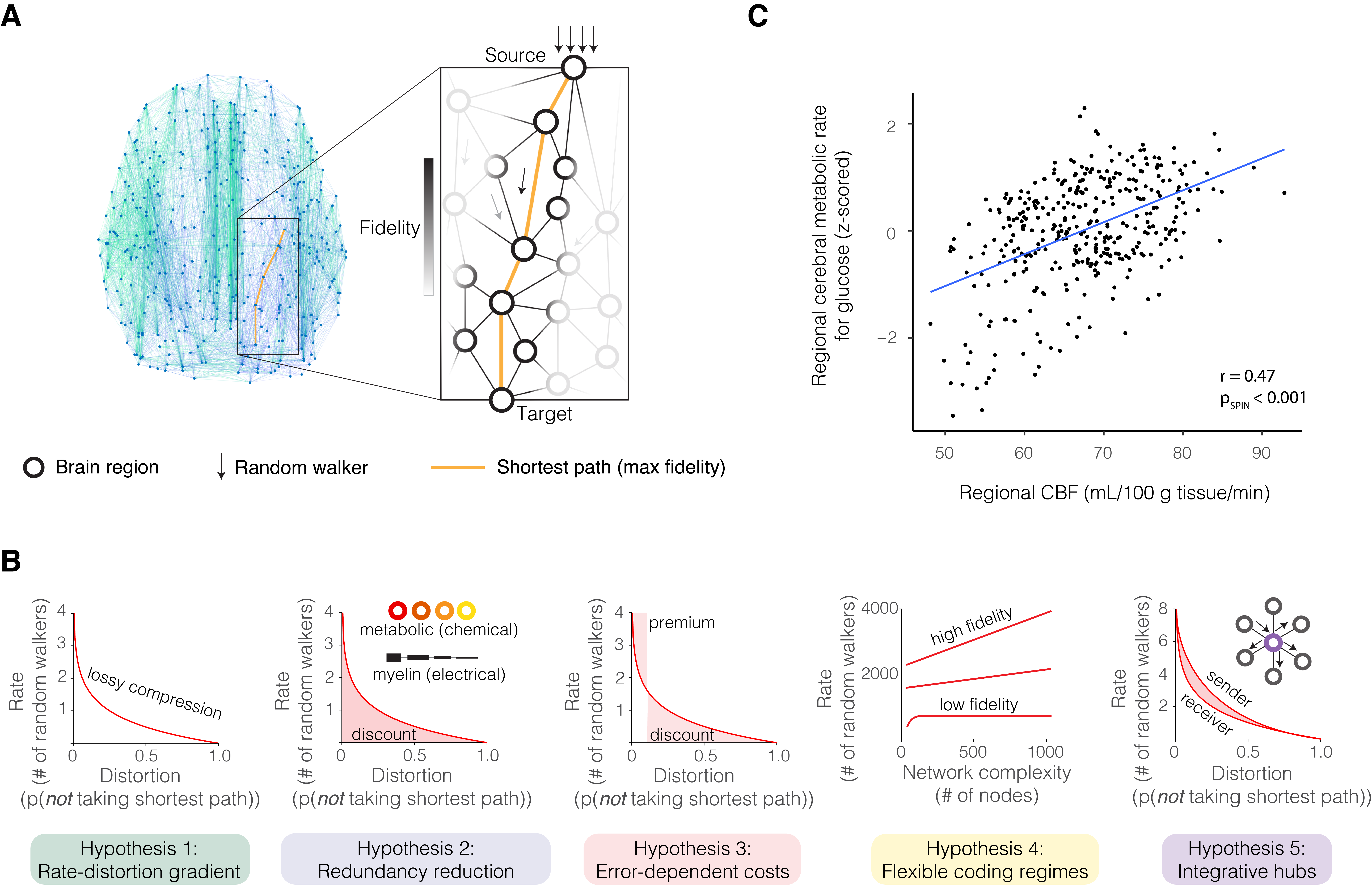}
    	\caption{\textbf{The efficient \rev{coding principle} and associated hypotheses.} \textbf{\emph{(A)}} To model efficient coding, we applied rate-distortion theory, a branch of information theory that provides the foundations of lossy compression \cite{shannon1959coding}, to a previously proposed network measure of random walk dynamics \cite{goni2013exploring}. This metric models random walks atop the structural connectome to solve for the number of random walkers required for at least one to propagate by the shortest path given an expected probability. After determining that the minimal amount of noise is achieved by signals that \rev{randomly walk} along shortest paths, we calculated the optimal information rate to communicate between brain regions with an expected transmission fidelity in the rate-limited (or capacity-limited) structural network. \textbf{\emph{(B)}} Across the brain, regional CBF measured from 1,042 participants in our study correlated with the regional cerebral metabolic rate for glucose acquired from published maps of 33 healthy adults (Pearson's correlation coefficient $r=0.47, df=358, p_\textrm{SPIN}<0.001)$, replicating prior findings and supporting the operationalization of metabolic expenditure using CBF \cite{Vaishnavi17757, gur2008regional}. We used CBF to investigate the relationship between metabolic demands and network organization supporting either shortest path routing (a previously posited communication mechanism) or random walks (our posited communication mechanism). \textbf{\emph{(C)}} Importantly, our model generates five predictions from the literatures of rate-distortion theory and neural information processing \cite{sims2018efficient, marzen2017evolution, van2012high}. 
		\label{fig1}}
	\end{figure}
    
    \rev{A parsimonious model of information transmission in connectomes emerges naturally from two key assumptions. The first assumption is that stochastic transmission entails an economy of discrete impulses where the immediate future state only depends on the current state \cite{barlow1961possible, 10.2307/j.ctt17kk982}. Mathematically, this assumption casts transmission as a linear process, or a random walk, wherein message copies (identical random walkers) propagate along structural connections with probabilities proportional to the connection's microstructural integrity \cite{2016207, avena2018communication}. Biologically, macroscale random walk models are supported by their ability to predict the transsynaptic spread of pathogens \cite{raj2015network, zheng2019local, henderson2019spread}, as well as the directionality and spatial distribution of neural dynamics from structural connectivity \cite{goni2014resting, abdelnour2018functional, seguin2019inferring, paquola2020cortical}. The second assumption is that the impulse can lose information but never generate additional information over successive steps of propagation \cite{amico2019towards}. Mathematically, this assumption represents the data processing inequality, which states that a random walker can only lose (and never gain) information about an information source \cite{cover1999elements}. Biologically, the assumption is supported by increasing temporal delay, signal mixing, and signal decay introduced by longer paths \cite{murray2014hierarchy, 10.2307/j.ctt17kk982}.} \\
    
    \rev{Combining these two assumptions, if packets of information propagate along structural pathways and information can only be lost with each step, then the shortest pathway between two brain regions yields an upper bound on the fidelity with which they can communicate. This key conclusion allows one to formulate the probability that a message propagates along the shortest path as an effective fidelity for the communication between two regions, thereby operationalizing the notion of \textit{distortion}. Moreover, by modeling messages as random walkers, one can operationalize the notion of \textit{rate} by computing the number of messages that must be sent to ensure that at least one transmits along the shortest path; that is, to ensure that at least one message reaches a specified receiver with maximum fidelity \cite{goni2013exploring}.} We applied our model to 1,042 youth (aged 8-23 years) from the Philadelphia Neurodevelopmental Cohort who underwent diffusion \rev{weighted} imaging (DWI; see \textbf{Supplementary Figure 1}) \cite{satterthwaite2014neuroimaging}. To operationalize metabolic expenditure, we used arterial-spin labeling (ASL) MRI, which measures cerebral blood flow (CBF) and is correlated with glucose expenditure \cite{Vaishnavi17757, gur2008regional}. \\
    
    \rev{To evaluate the validity of the efficient coding model, we assessed five published predictions of any communication system adhering to rate-distortion theory, which we adapted to connectomes and distinguished from alternative explanations of brain network communication dynamics (\textbf{Figure 1B}) \cite{van2012high, goni2013exploring, goni2014resting,  marzen2017evolution, sims2018efficient}. First, information transmission should produce a characteristic rate-distortion gradient in biological and artificial networks, where exponentially increasing information rates are required to minimize signal distortion. Second, transmission efficiency should improve with manipulations of the communication system designed to facilitate signal propagation, where information costs decrease when \rev{randomly walking messages are} biased with regional differences in metabolic rates \rev{and intracortical myelin}. Third, the information rate should vary as a function of the costs of error, with discounts when costs are low and premiums when costs are high. Fourth, brain network complexity should flexibly support communication regimes of varying fidelity, where a high-fidelity regime predicts information rates that monotonically increase as the network grows more complex, and a low-fidelity regime predicts asymptotic information rates indicative of lossy compression. Fifth and finally, structural hubs should integrate incoming signals to efficiently broadcast information, where hubs (compared to other brain regions) have more compressed input rates and higher transmission rates for equivalent input-output fidelity. As described below, this model advances the current understanding of how information processing is associated with behaviors in a range of cognitive domains, subject to constraints on metabolic resources and network architecture.}
	
	\section*{Results}
	
	\subsection*{Macroscale efficient coding can be understood by communication processes of random walks but not the alternative model of shortest path routing}
	
\begin{figure}[htbp!]
	\centering
	\includegraphics[width=1 \columnwidth]{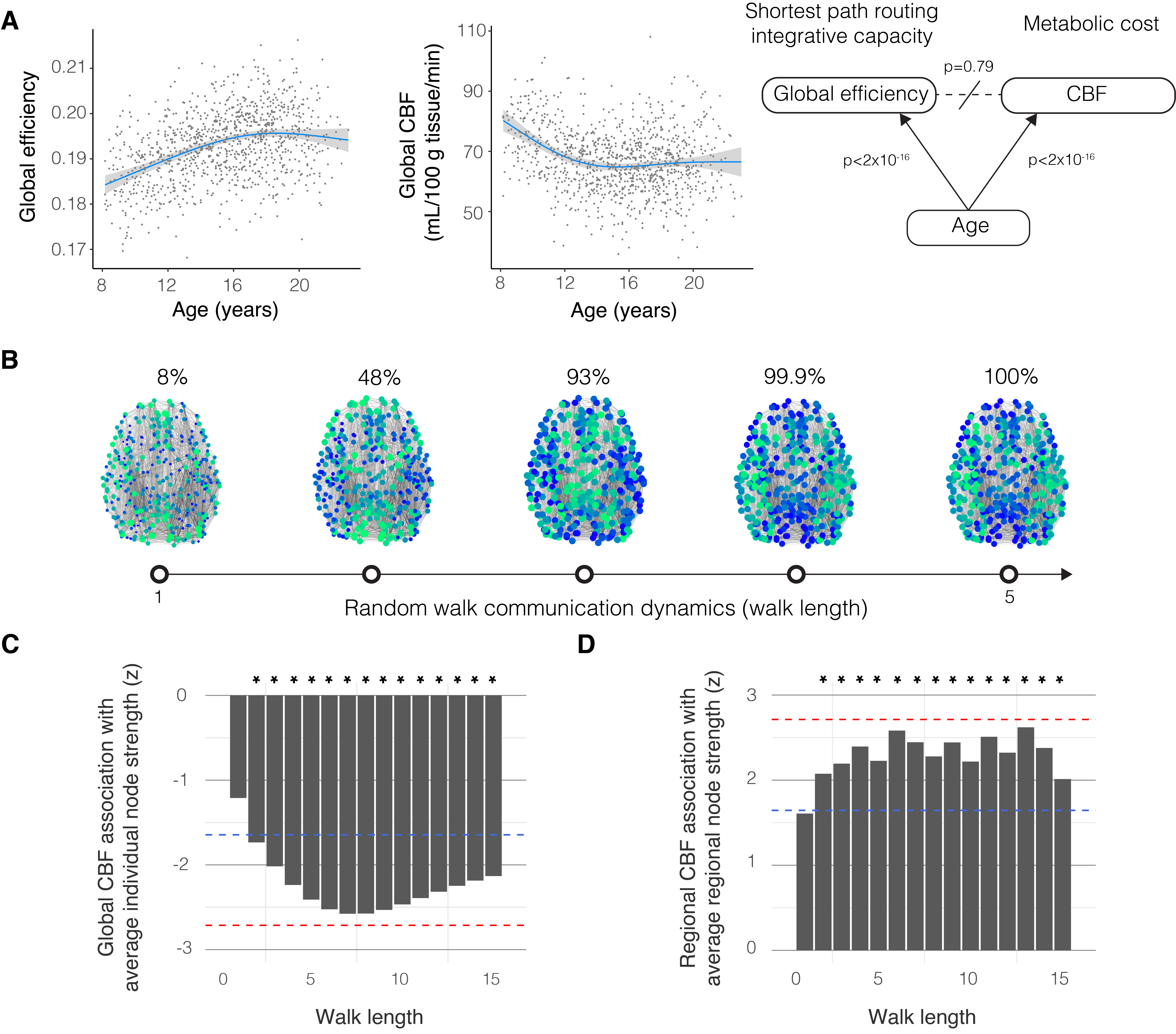}
	\caption{\textbf{Metabolic running costs support brain network architectures for random walk dynamics, not shortest path routing.} \textbf{\emph{(A)}} Global efficiency increased with development while cerebral perfusion declined ($F_{\textrm{global efficiency, age}}=50$, estimated $df=3.46$, $p<2 \times 10^{-16}$; $F_{\textrm{CBF,age}}=69.22$, estimated $df=3.74$, $p<2\times10^{-16}$; confidence intervals displayed as gray bands). Contrary to prior reports, the relationship between CBF and global efficiency was confounded by age ($r=0.01, df=1039, p=0.79$). Therefore, the claim of reduced metabolic cost associated with shortest path routing is weakened. \textbf{\emph{(B)}} All brain regions are accessible to signals randomly walking across 5 edges or more. Bluer nodes represent greater normalized node strengths, reflecting the accessibility of the brain region to a random walker propagating along the structural connectome with differing walk lengths. Whereas the shortest path routing model predicts that only shortest path walks of length 5 or less are associated with CBF, the random walk model predicts that longer random walk paths are also associated with CBF. \textbf{\emph{(C)}} Global CBF was negatively correlated with the strengths of structural paths with length 2 or more when controlling for age, sex, age-by-sex interactions, degree, density, and in-scanner motion ($t=-1.59$ to $-2.81$, estimated model $df=11.45$). Asterisks denote statistical significance following correction for multiple comparisons by FDR; blue dashed line: FDR-corrected $p = 0.05$, red dashed line: Bonferroni-corrected $p = 0.05$ for 15 tests. Metabolic expenditure was related to structure supporting random walks rather than shortest path routing. \textbf{\emph{(D)}} Regional CBF was positively correlated with high-integrity connections comprising the structural scaffolds supporting differing walk lengths ($\rho=0.12$ to $0.14$, $df=358$, FDR-corrected $p < 0.05$), controlling for age, sex, age-by-sex interaction, degree, density, and in-scanner motion. Blue dashed line: FDR-corrected $p = 0.05$, red dashed line: Bonferroni-corrected $p = 0.05$ for 15 tests. Together, these data provided some evidence for metabolic cost associated with a regional profile of white matter path strengths that support random walk dynamics. Individuals with greater path strengths tended to have lower global metabolic expenditure, while brain regions with greater path strengths tended to have greater metabolic expenditure.
		\label{fig2}}
\end{figure}

    To understand how the brain balances the transmission rate of \rev{stochastic messages} and signal distortion across different network architectures, we developed a model positing random walk communication dynamics atop the structural connectome. \rev{\rev{We tested this model by comparing it to the alternative hypothesis of shortest-path routing (see \textbf{Supplementary Modeling/Math Notes}).}} The random walk model and shortest path model are currently viewed as opposing extremes of a spectrum of communication processes \cite{goni2013exploring, avena2015network, avena2018communication}. In network neuroscience, shortest path routing anchors metrics of communication dynamics and information integration \cite{sporns2013network, Seguin6297, avena2018communication}. In cognitive neuroscience, the neural circuit related to behavior is commonly depicted as a subset of brain regions communicating their specialized information to each other across shortest and direct anatomical connections \cite{10.3389/fnins.2019.00897}. Although shortest path routing has acknowledged shortcomings as a model of communication dynamics (see \textbf{Supplementary Modeling/Math Notes}), a key extenuating hypothesis of the model is reduced metabolic cost \cite{bullmore2012economy, avena2018communication}. Yet, existing macroscale evidence for this model remains sparse \cite{varkuti2011quantifying}. \\
    
    We sought to determine how brain metabolism is associated with structural signatures of shortest path versus random walk models. To quantify the extent to which a person's brain is structured to support shortest path routing, we used a statistical quantity known as the \emph{global efficiency} \cite{latora2001efficient}, a commonly used measure of the average shortest path strength between all pairs of brain regions. Intuitively, global efficiency represents the ease of routing information by shortest paths and is proportional to the strength of shortest paths in a network \cite{sporns2013network}. As an operationalization of metabolic running cost, we considered CBF, which is correlated with glucose consumption \cite{Vaishnavi17757, gur2008regional}. To test the spatial correlation between CBF and glucose consumption, we used a spatial permutation test that generates a null distribution of randomly rotated brain maps that preserves the spatial covariance structure of the original data; we denote the $p$-value that reflects significance as $p_{\textrm{SPIN}}$ (Method 7.7.6). We observed a linear association between CBF and glucose consumption (\textbf{Figure 1C}; Pearson's correlation coefficient $r=0.47, df=358$, $p_{\textrm{SPIN}}<0.001$). \\
    
    Next, we tested if shortest path routing is linked to a decrease in metabolic expenditure, operationalized as a negative correlation between global efficiency and CBF \cite{varkuti2011quantifying, bullmore2012economy}. Controlling for mean gray matter density, sex, mean degree, network density, and in-scanner motion, we found that the global efficiency was negatively correlated with CBF ($r=-0.20, df=1039, p<0.001$), consistent with prior reports \cite{varkuti2011quantifying}. Notably, we did not regress out age in the previous analysis in order to align with the prior analysis that we aimed to replicate \cite{varkuti2011quantifying}. We were also interested in determining whether development had any effect on the relationship between global efficiency and CBF. Our interest was justified by the positive correlation between age and global efficiency (\textbf{Figure \ref{fig2}A}; $F=50$, estimated $df=3.46$, $p < 2\times10^{-16}$) and the negative correlation between age and CBF ($F=69.22$, estimated $df=3.74$, $p<2\times10^{-16}$). After controlling for age we did not find a significant relationship between global efficiency and CBF ($r = 0.01, df=1039, p=0.79$), suggesting that co-linearity with age drove the initial observed association between CBF and global efficiency. This null result undermines the claim that shortest path routing is associated with reduced metabolic expenditure.\\
    
    Rather than being driven by shortest path routing, metabolic expenditure could instead be associated with communication by random walks. Each brain region can reach every other brain region via random walks along paths of 5 connections (\textbf{Figure \ref{fig2}B}). A random walker will likely not take the most efficient paths and must instead rely on the structural strengths of longer paths. Hence, if brain metabolism is associated with communication by random walks, then CBF should correlate with the strength of the white matter paths greater than length 5. To evaluate this prediction, we computed the strength of connections across different path distances using the matrix exponent of the structural network (see Method \ref{pathStrengths} and \textbf{Supplementary Figure 1B}). We then tested the association between longer paths and metabolic expenditure across individuals (\textbf{Figure \ref{fig2}C}) and across regions (\textbf{Figure \ref{fig2}D}). In first considering variation across \textit{individuals}, we found that the average node strengths for walks of length 2 to 15 were negatively correlated with CBF ($t=-1.59$ to $-2.81$, estimated model $df=11.45$, FDR-corrected $p<0.05$), after controlling for age, sex, age-by-sex interaction, average node degree, network density, and in-scanner motion (\textbf{Figure \ref{fig2}C}). The negative correlations between CBF and the average connection strengths suggest that the greater the connection integrity, the lower the metabolic expenditure. In next considering variation across \textit{brain regions}, we found that the average node strengths for walks of length 2 to 15 were positively correlated with CBF (Spearman's rank correlation coefficient $\rho=0.12$ to $0.14$, $df=358$, FDR-corrected $p<0.05$), after controlling for age, sex, age-by-sex interaction, average node degree, network density, and in-scanner motion (\textbf{Figure \ref{fig2}D}). The positive correlations between CBF and the average connection strengths suggest that brain regions with greater path strengths tended to have higher metabolic expenditure. See \textbf{Supplementary Figures 2-4} for metabolic costs associated with other connectivity metrics supporting random walks. Together, we found no evidence of metabolic expenditure associated with shortest path routing, whereas the convergent findings of an association between CBF and random walk path strengths across individuals and regions provided some evidence that metabolic running costs were linked to random walk communication dynamics. \\

	\rev{Using the random walk model, we formalized a rate-distortion model of efficient \rev{coding} by assuming that the minimal amount of noise is achieved by messages that \rev{randomly walk} along shortest paths (\textbf{Figure \ref{fig4}A-D}). \rev{We defined rate as the number of random walkers per transmission, and distortion as} the probability of \rev{the random walkers} \emph{not} taking the shortest path. We evaluated the validity of the redundancy reduction implementation of efficient coding \cite{barlow1961possible}. Reducing redundancy in repetition coding is equivalent to minimizing the number of random walkers \cite{barlow1961possible, shannon1959coding, cover1999elements}.} To understand how the brain balances information rate and distortion, we measured the number of random walkers that are required for at least one to randomly walk along the shortest path to a target cortical region, with an expected probability (Method \ref{resourceEfficiency}). \rev{This measure of random walk dynamics is based on a prior metric \cite{goni2013exploring}. The number of random walkers can be used to calculate the transmission length of a neural message (in units of bits) or the information rate (in units of bits per second; see Method \ref{shannonInfo} and \textbf{Supplementary Figure 3}).} To evaluate the roles of random walk dynamics and rate-distortion theory in the brain, we assessed five previously published predictions of rate-distortion theory and information \rev{processing} (\textbf{Figure \ref{fig1}C}) \cite{marzen2017evolution, sims2018efficient, van2012high}. 
	
	\subsection*{\rev{Hypothesis 1: Rate-distortion gradient}}
	
	\begin{figure}[htbp!]
	\centering
	\includegraphics[width=1\columnwidth]{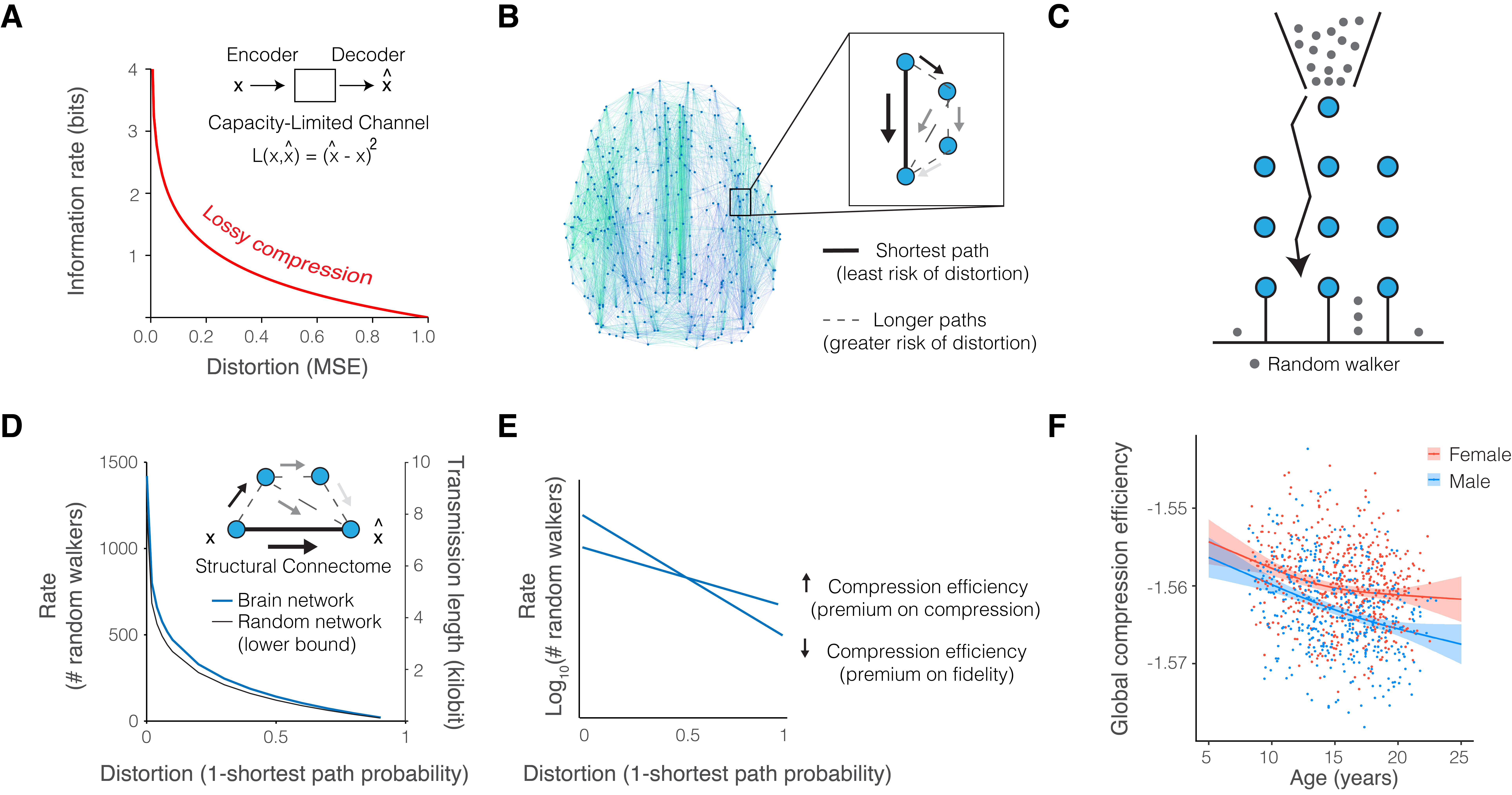}
        \caption{\textbf{\rev{Lossy compression supports efficient coding in brain networks.}} \textbf{\emph{(A)}} Rate-distortion theory is a mathematical framework that defines the required amount of information for an expected level of signal distortion during communication through capacity-limited channels. Loss functions, such as the mean squared error, are defined to map encoded and decoded signals. In a given channel, compressed messages demand lower information rates at the cost of fidelity. \textbf{\emph{(B)}} In brain networks, we defined the distortion function as the probability of a \rev{randomly walking} signal \emph{not} taking the shortest path. This definition is reasonable because the temporal delay, signal mixing, and decay introduced by longer paths collectively increase distortion. \textbf{\emph{(C)}} The Galton board depicts the problem of determining the minimum number of (\rev{random walkers}) that the starting node should prepare to transmit in order for one resource (random walker) to propagate by the specified path given an expected probability. Formalizing the shortest path distortion function allows analytical solutions to determine the corresponding minimum rate of information using an established metric \cite{goni2013exploring}. This metric provided us with the number of random walkers (or amount of information) required for at least one walker to propagate by the shortest path with an expected probability. \textbf{\emph{(D)}} In applying our theory to the structural connectome, we observed a characteristic curve that was consistent with that predicted by rate-distortion theory. The artificial random networks exhibited the rate-distortion gradient also observed in brain networks, supporting the first prediction of rate-distortion theory that both biological and artifical communication systems are governed by the same information-theoretic trade-offs. The blue curve depicts the locally estimated scatterplot smoothing fit of the mean rate-distortion gradient across all individuals. The black curve depicts the mean rate-distortion gradient for Erd\H{o}s-Reny\'{i} random networks whose edges maintain the weights from empirical measurements. The number of random walkers required of brain connectomes was greater than the number required of random networks, across levels of distortion ($F=9.6 \times 10^5, df=29120$, \rev{bootstrap 95\% CI [$9.4 \times 10^5, 10.4 \times 10^5$]}, $p<2 \times 10^{-16}$, \rev{adjusted $R^2$=99.9\%, bootstrap 95\% CI [0.9989, 0.999]}), highlighting the established graph theoretical notion that random networks increase shortest path accessibility but demand a biologically unrealistic cost of materials for structural connections \cite{goni2013exploring}. \rev{The information content of a random walker can be calculated in order to interchange the number of random walkers for the transmission length of a neural message in units of bits (see Method \ref{shannonInfo} and \textbf{Supplementary Figure 3}).} \textbf{\emph{(E)}} When plotted on a logarithmic axis, the exponential rate-distortion function appears as a straight line. We measured the individually different slopes as differing \emph{compression efficiency}. Then, we interpreted variation across individuals in the language of rate-distortion theory. For example, consider two brain networks functioning at the same low level of distortion. The brain network with the flatter slope between random walkers and distortion has greater compression efficiency because the network architecture confers information rate discounts. In comparison, the brain network with a steeper slope has reduced compression efficiency because the network architecture pays a premium for the same expected fidelity. \textbf{\emph{(F)}} When we considered variation across individuals, we found that compression efficiency decreased with age and differed by sex (\textbf{Supplementary Figure 5}; $F=27.54$, \rev{bootstrap 95\% CI [1.78, 97.52]}, estimated $df=2.17$, $p<0.001$; \rev{adjusted $R^2$=17.2\%, bootstrap 95\% CI [0.13, 0.23]}; confidence intervals displayed as the colored bands), suggesting that neurodevelopment places a premium on high-fidelity network communication.
	\label{fig4}}
	\end{figure}
	
	The first prediction of rate-distortion theory is that communication systems including both brain networks and artificial random networks should produce an information rate that is an exponential function of distortion because biological and engineered systems are governed by the same information-theoretic trade-offs \cite{sims2018efficient}. To test this prediction, we computed the \rev{average number of random walkers over all nodes in the network for a given individual}, with the probability of \rev{randomly walking} along the shortest path ranging from 10\% to 99.9\% (\textbf{Figure \ref{fig4}D}). We compared the number of random walkers required of structural connections in the brain with that of random walkers required of connections in random Erd\H{o}s-Reny\'i networks, which have larger probabilities of shortest path communication \revtwo{compared amongst canonical random networks} \cite{latora2001efficient, goni2013exploring}. \revtwo{Hence, Erd\H{o}s-Reny\'{i} networks serve as an optimal benchmark for the efficiency of random walk communication and shortest path routing  (Method \ref{networkNulls}).} For each individual network, the information-theoretic trade-off between information rate and signal distortion was defined by a rate-distortion gradient (\textbf{Figure \ref{fig4}E}). The gradient shows that distortion increases as the information rate decreases, which is the hallmark feature of lossy compression. Next, we considered the extent to which the brain's structural connectome prioritizes compression versus fidelity. We refer to this trade-off as the \emph{compression efficiency} (\textbf{Figure \ref{fig4}E}), and define it as the slope of the rate-distortion gradient (Method \ref{compressionEfficiency}). With random walk communication dynamics, increased compression efficiency prioritizes lossy compression, while decreased compression efficiency prioritizes transmission fidelity. \\
	
	Consistent with the first prediction of rate-distortion theory, we observed an exponential gradient in every individual brain network and the Erd\H{o}s-Reny\'i random networks. Furthermore, the random networks, which are composed of more short connections than empirical brain networks, required significantly fewer random walkers than the empirical brain networks (\textbf{Figure \ref{fig4}D} and \textbf{Supplementary Figure 3}; $F=10\times10^{5}, df=29120, p<2\times10^{-16}$), consistent with the intuition that a greater prevalence of short connections in the random network translates to greater likelihood of shortest path \rev{propagation} \cite{latora2001efficient, goni2013exploring}. Rate-distortion trade-offs varied as a function of age and sex, where compression efficiency (\textbf{Figure \ref{fig4}E}) was negatively correlated with age ($F=27.54$, estimated $df=2.17$, $p<0.001$), suggesting that neurodevelopment places a premium on fidelity (\textbf{Figure \ref{fig4}F}). Compression efficiency was greater on average in females compared to males ($t=9.53$, $df=996.82$, $p<0.001$). The data, therefore, indicate that random walk communication dynamics on biological brain networks differ from random walks on artificial networks, yet each accords well with the prediction of rate-distortion trade-offs governing all communication systems.
	
	\subsection*{\rev{Hypothesis 2: Redundancy reduction}}
	
	\begin{figure}[htbp!]
	\centering
	\includegraphics[width=1\columnwidth]{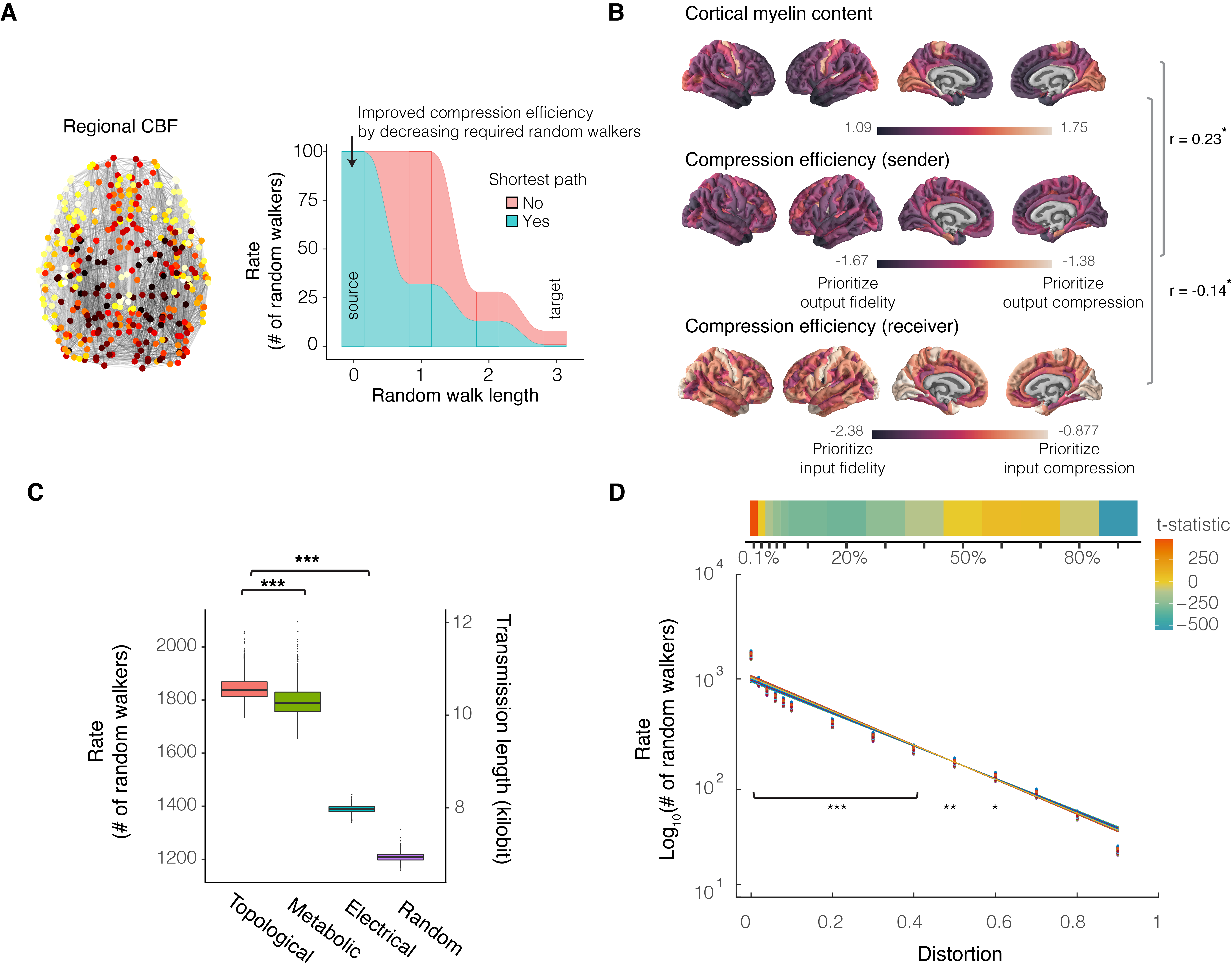}
		\caption[short]{\textbf{\rev{Connectome architecture and biology support efficient coding.}} \textbf{\emph{(A)}} Biased random walks with transition probabilities modified by regional CBF \rev{and myelin content are biologically motivated by two fundamental modes of neural communication: biochemical diffusion and electrical signaling.} Left: A biased random walk attracting \rev{walkers} to regions of high CBF was modeled by constructing a biased structural connectivity matrix. The weight of an edge between region $i$ and region $j$ was given by the average, normalized CBF between pairs of brain regions multiplied by the weight of the structural connection. Right: The alluvial diagram depicts our expected results: if \rev{CBF} acts as a substrate supporting efficient coding, then we should observe reduced minimum random walkers in \rev{random walks biased by CBF} compared to \rev{random walks propagating only based on connectome topology}. \textbf{\emph{(B)}} \rev{Sender and receiver compression efficiency differ regionally across the cortex and describe the number of \rev{randomly walking} messages required to leave from or arrive to a brain region with specified signal fidelity (\textbf{Supplementary Figure 6}). Regional values were averaged across individuals. Brain regions differ in myelination according to published maps \cite{glasser2014trends}. Cortical regions with greater levels of myelin tended to have greater sender compression efficiency ($r=0.23$, $p_{\textrm{SPIN, Holm-Bonferroni}}=0.02$) and receiver compression efficiency ($r=0.14$, $p_{\textrm{SPIN, Holm-Bonferroni}}=0.046$), consistent with myelin reducing conduction delay and promoting the efficient trade-off between signal rate and fidelity to reduce the transmission rate while preserving fidelity (see \textbf{Supplementary Figure 7}). \textbf{\emph{(C)}} This plot depicts the number of random walkers required to communicate with 0.01\% distortion (see \textbf{Supplementary Figure 6C} for distortion ranging from 2\% to 90\%). In agreement with our second hypothesis that \rev{connectome structure and biology} supports efficient coding, the number of required random walkers decreased in \rev{random walks biased by CBF (metabolic)} compared to \rev{unbiased random walks only based on connectome topology (topological) when the level of distortion was fixed at 0.01\%} \rev{($t_{\textrm{topology,metabolic}}=20.87$, $\textrm{bootstrap 95\% CI} [18.72, 23.14]$, $df=1993.9$). Similarly, fewer random walkers were required to communicate in the intracortical myelin-biased random walk (electrical) compared to the unbiased random walk $t_{\textrm{topology,electrical}}=295.93$, $\textrm{bootstrap 95\% CI} [281.19, 312.76]$, $df=1225.9$). Random walks on random networks required the least random walkers consistent with the notion that random networks increase shortest path }}
			\label{fig5}}
	\end{figure}
	
	\begin{figure}[htbp!]
		\ContinuedFloat
		\caption[]{\textbf{(continued.)} \rev{accessibility but demand a greater materials cost ($t_{\textrm{topology,random}}=405.39$, $\textrm{bootstrap 95\% CI} [386.3, 427.70]$, $df=1307.3$; ***: $p<0.001$).} \textbf{\emph{(D)}} Here, we depict rate-distortion gradients per individual structural connectome for communicating across a range of transmission fidelity (1-distortion). When plotted on a logarithmic axis, the exponential functions appear linear. The level of distortion and type of random walk fully explained variance in the number of random walkers (\rev{ $F=8.1 \times 10^5$, bootstrap 95\% CI [$7.8 \times 10^5, 8.5 \times 10^5$], df=58240, adjusted $R^2$=99.9\%, bootstrap 95\% CI [0.9986, 0.9987]}; all $p$-values corrected using the Holm-Bonferroni method for family-wise error rate; ∗ : $p < 0.05$, ∗∗ : $p < 0.01$, ∗∗∗ : $p < 0.001$). According to the third prediction of rate-distortion theory, we should expect asymmetries about the rate-distortion gradient where there are asymmetric costs of error. Specifically, the cost of error should be greater during neural dynamics requiring high-fidelity communication compared to low-fidelity communication. Across individual brain networks, the random walkers required for 0.1\% distortion were greater than predicted by the rate-distortion gradient, reflecting a premium placed on very high fidelity signaling. \rev{At levels of distortion greater than 2\% and less than 60\%, the random walkers required were less than that predicted by the rate-distortion gradient.} The premium cost of high-fidelity communication compared to discounts of low-fidelity communication supports the third prediction of rate-distortion theory. }
	\end{figure}	
	
	\rev{The second prediction of rate-distortion theory is that manipulations to the physical communication system (the connectome) that are designed to facilitate information transmission will improve communication efficiency. The prediction stems from two key observations. First, between the rate-distortion gradients for structural connectomes and random networks (depicted in \textbf{Figure \ref{fig4}D}) exists a range of possible rate-distortion gradients produced by some other construction of communication networks \cite{shannon1959coding}.} \rev{Second, it is well known that brain signaling relies extensively on metabolic diffusion and devotes much of its metabolic resources to maintaining a chemical balance that supports neuron firing \cite{10.2307/j.ctt17kk982, Attwell2001AnEB}. At the longer distances of the connectome, myelin in the white matter and cerebral cortex supports the speed and efficiency of electrical signaling in subcortical fiber tracts and in cortico-cortical communication \cite{barbas1997cortical, laughlin2001energy, deco2014cortico}.} \rev{Hence, modifying the connectome to bias random walk dynamics according to metabolic resources and myelin mimics biological investments in communication efficiency.}\\
	
	\rev{The second observation above leads to the hypothesis that including biological biases in random walk probabilities based on the strength of structural connections will improve the efficiency of information transmission. We hypothesized that biasing random walkers with metabolic resources and myelin would reduce the information rate required to communicate a message with a given fidelity compared to random walkers that propagate only by connectome topology.} \rev{To test this hypothesis, we biased edge weights (Method \ref{biasedRW}) representing structural connection strength by multiplying the edge weight by a bias term. Across pairs of connected brain regions, the bias term was either defined as the average, normalized metabolic rate using CBF or the average, normalized cortical myelin content using published maps of T2/T1w MRI measures with histological validation (\textbf{Figure \ref{fig5}A}) \cite{glasser2014trends}. By modeling network communication dynamics, one can calculate directed patterns of transmission as inputs into (receiver) and outputs from (sender) brain regions \cite{seguin2019inferring}. We separately computed the send and receive compression efficiency of brain regions to better understand the biological relevance of transmitted information sent or received across the connectome (\textbf{Figure \ref{fig5}B}; Method \ref{compressionEfficiency}).} \\
	
	\rev{We found that brain regions that prioritized input fidelity and output compression tended to have greater myelin content (\textbf{Figure \ref{fig5}B}; sender $r=0.23$, $df=358$, $p_{\textrm{SPIN, Holm-Bonferroni}}=0.02$, receiver $r=-0.14$, $df=358$, $p_{\textrm{SPIN, Holm-Bonferroni}}=0.046$), consistent with myelin's function in neurotransmission efficiency and speed. For a channel communicating at 0.1\% distortion (and across all distortions; \textbf{Supplementary Figure 6C}), biasing structural edge weights by the biological properties of metabolic and electrical signaling resulted in more efficient communication by reducing the number of redundant random walkers required (\textbf{Figure \ref{fig5}C}; $t_{metabolic}=20.87$, $\textrm{bootstrap 95\% CI}$ $[18.72, 23.14]$, $df=1993.9$; $t_{electrical}=295.93$, $\textrm{bootstrap 95\% CI}$ $[281.19, 312.76]$, $df=1225.9$, $p<0.001$)}. \rev{While both metabolic and electrical signaling supported more efficient communication, electrical signaling was more efficient than metabolic signaling \cite{10.2307/j.ctt17kk982, deco2014cortico}.} \revtwo{Compared to rewired null networks preserving the degree sequence (\textbf{Supplementary Figure 6D}), structural topology and metabolic resources support communication that prioritizes fidelity ($t_{\textnormal{topological,degree-preserving}}(2074.4) = 121.02, p < 0.001$; $t_{\textnormal{metabolic,degree-preserving}}(2025.8) = 87.78, p < 0.001$), while myelination supports communication that prioritizes compression efficiency ($t_{\textrm{electrical,}}$ $_\textrm{degree-preserving}$ (1208.3) = -122.62, $p < 0.001$)}. The minimum number of random walkers required for distortion levels less than 60\% was explained by the interaction of the distortion level with the type of \rev{biased random walk} (see \textbf{Figure \ref{fig5}C}, $F=6\times10^{5}, df=29120$, $p_{\textrm{Holm-Bonferroni}}<0.05$). Together, these results support the prediction that \rev{biological investments can augment connectivity to support efficient communication, especially when transmission prioritizes fidelity.}
	
	\subsection*{\rev{Hypothesis 3: Error-dependent costs}}
	
	The third prediction of rate-distortion theory is that the information rate should vary as a function of the costs of errors in communication systems that interact with their environment. If errors are more costly for networks operating at high fidelity, then we should observe an information rate surpassing the minimum predicted by rate-distortion theory. In contrast, if errors are less costly for networks operating at low fidelity, then we should observe no more than the minimum predicted information rate. \rev{In testing this prediction, w}e observed that brain networks commit more random walkers than required for very low levels of distortion, such as 0.1\%, but allocate the predicted number of random walkers or fewer to guarantee levels of distortion between 2\% and \rev{60\%} (\textbf{Figure \ref{fig5}D}). Hence, the third prediction of rate-distortion theory was consistent with our observation of a premium placed on very low signal distortion and a discounted cost of greater distortion. 
		
	\subsection*{\rev{Hypothesis 4: Flexible coding regimes}}
	
	The fourth prediction of \rev{rate-distortion theory proposes that communication systems, including the connectome, have distinct network properties that support information transmission in a flexibly high- or low-fidelity regime. With increasing information processing demands, a high-fidelity regime will continue to place a premium on accuracy, whereas a low-fidelity regime will tolerate noise in support of lossy compression. The operating regime depends on the behavioral demands of the environment, indicating the need for a flexible regime that can simultaneously support both high- and low-fidelity communication. In this section, we assess the fourth prediction of rate-distortion theory by testing the more precise hypotheses that large brain networks support communication in a high-fidelity regime, indirect pathways supports a low-fidelity regime, and hierarchical organization supports a flexible regime. We explain and test each hypothesis in turn.}
	
	\subsubsection*{Large networks support a high-fidelity regime and indirect pathways support a low-fidelity regime}
	
	Rate-distortion theory predicts that, in a high-fidelity regime, the information rate will monotonically increase with the complexity of the communication system in order to continue to place a premium on accuracy (\textbf{Figure \ref{fig6}A}). To evaluate this prediction, we operationalized complexity as network size because size 
	
	\begin{figure}[H]
	\centering
	\includegraphics[width=1\columnwidth]{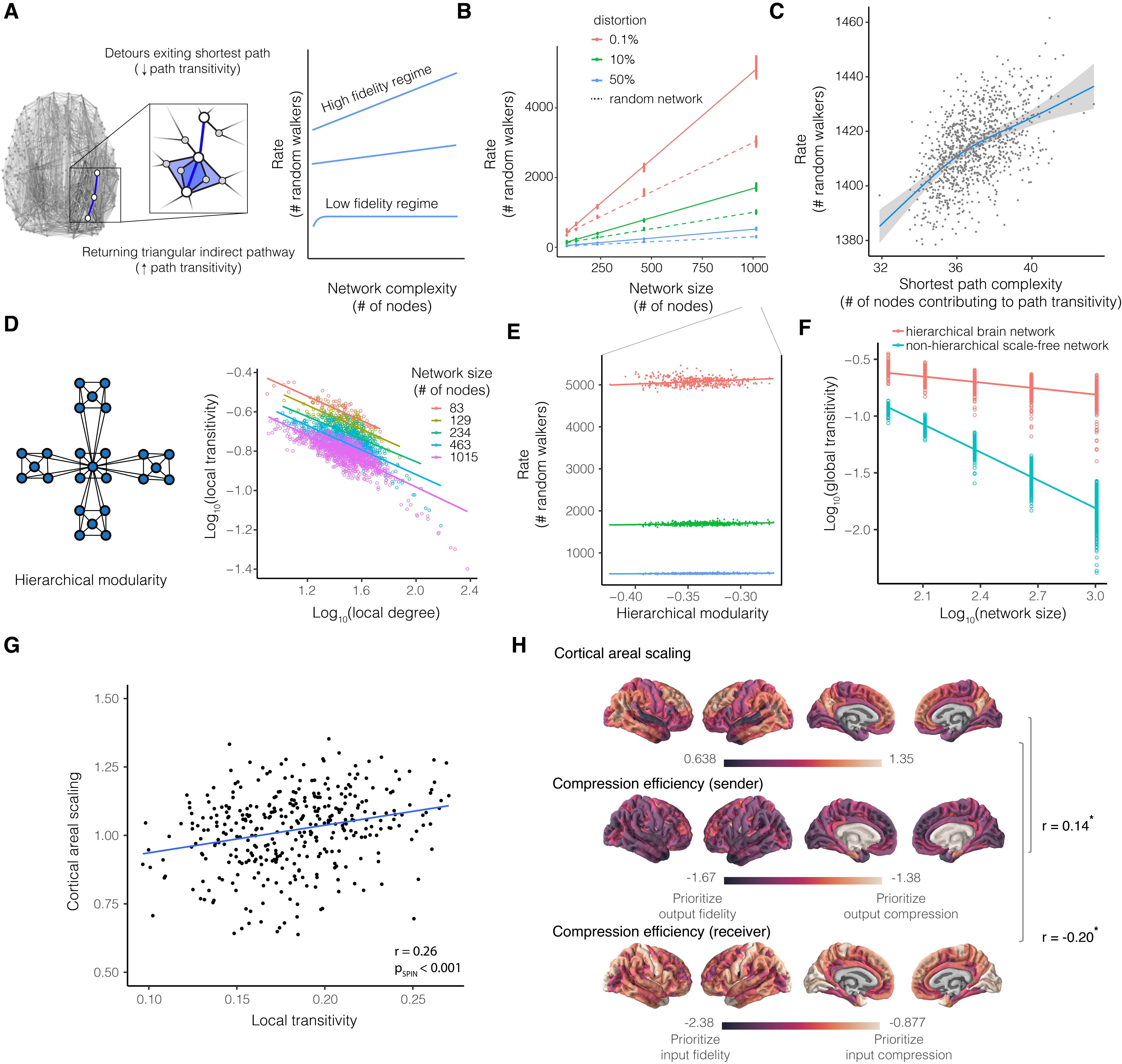}
		\caption[short]{\textbf{\rev{Constraints of compression efficiency and }brain network complexity.} \textbf{\emph{(A)}} High versus low fidelity communication regimes make different predictions regarding the minimum number of random walkers required for an expected distortion and system complexity. We define system complexity as the network size, or the number of functionally and cytoarchitectonically distinct brain regions. In the high fidelity regime, the brain should place a premium on high fidelity signaling by allocating a monotonically increasing information rate to achieve low distortion despite increasing system complexity. In the low fidelity regime, the brain should place a premium on lossy compression by asymptotically capping the information rate despite increasing system complexity. We expected the number of triangular pathways along the shortest path (i.e., the path transitivity) to approximate the shortest paths with a one-connection longer detour. The fewer the pathways approximating the shortest and highest fidelity pathway, the lower the fidelity of communication. \textbf{\emph{(B)}} Number of random walkers scaled monotonically with network size, consistent with the prediction of a high fidelity regime. Individual brain networks were reparcellated using the Lausanne brain atlases with 83, 129, 234, 463, and 1015 parcels, respectively. Together, the monotonically increasing resource demand highlights a trade-off between high fidelity communication and network complexity, indicating an additional evolutionary constraint on brain network size and complexity. \textbf{\emph{(C)}} Shortest path complexity is the average number of nodes comprising path transitivity. Number of random walkers scaled asymptotically with inter-individual differences in path transitivity, consistent with a low-fidelity regime for lossy compression and storage savings. The non-linear model outperformed a linear model (non-linear $AIC=7902$, linear $AIC=7915$; non-linear $BIC=7964$, linear $BIC=7968$; confidence intervals displayed as gray band). Lossy compression of a signal} 
		\label{fig6}
	\end{figure}
	
	\begin{figure}[H]
		\ContinuedFloat
		\caption[]{\textbf{continued}. by differing densities of path transitivity is akin to image compression by pixel resolution. Reduced path transitivity, or pixel resolution, tends to result in more compression, while greater path transitivity supports greater fidelity. \rev{\textit{\textbf{(D)}} Hierarchical modularity is a form of network complexity that is characterized by small groups of nodes organized successively into increasingly larger groups (left). A hallmark of hierarchical modularity is a strict scaling of the transitivity and degree, representing the clustering and scale-free property of nodes (right). Hierarchical organization was evident across the average reparcellated Lausanne brain networks. \textit{\textbf{(E)}} Hierarchical organization was monotonically related to increasing resource demands, but to a lesser cost compared to network size. Hence, hierarchical modularity may support both a high- and low-fidelity regime by increasingly balancing global and local network complexity. \textit{\textbf{(F)}} Hierarchically modular brain networks allowed for greater transitivity (clustering) despite increasing network sizes compared to scale-free networks that do not have a hierarchical architecture. \textit{\textbf{(G)}} Cortical areal scaling in neurodevelopment reflects patterns of evolutionary remodeling. Brain regions with greater transitivity, which supports greater fidelity, tended to disproportionately expand in relation to total brain size during neurodevelopment.} \textit{\textbf{(H)}} Brain regions that have higher sender compression efficiency tended to disproportionately expand in relation to total brain size during neurodevelopment ($r=0.14$, $df=358$, $p_{\textrm{SPIN, Holm-Bonferroni}}=0.046$). Cortical regions with low receiver compression efficiency, placing a premium on information processing fidelity, tended to disproportionately expand in relation to whole brain growth ($r=-0.20$, $df=358$, $p_{\textrm{SPIN, Holm-Bonferroni}}=0.03$). Positively scaling regions that prioritize compression-efficient broadcasting of messages arriving with high fidelity may reflect evolutionary expansion of brain regions with high information processing capacity, whereas negatively scaling regions that prioritize high-fidelity broadcasting of compressed messages may permit other modes of material, spatial, and metabolic cost efficiency.}
	\end{figure}
	
	\rev{determines the number of possible states (or nodes) available to each random walker }\cite{marzen2017evolution}. We re-parcellated each individual brain network at different spatial resolutions to generate brain and random null networks of 5 different sizes representing the original network. We compared the rate for brain networks to the rate for random networks of matched size. In testing this prediction, we observed that the minimum number of random walkers increased monotonically with network size, consistent with a high-fidelity regime, and at a rate different to random networks with matched sizes (\textbf{Figure \ref{fig6}B}). Larger brain networks support high fidelity communication by placing a greater premium on accuracy than do larger random networks.\\
	
	In addition to high-fidelity communication, in a low-fidelity regime, rate-distortion theory predicts that the information rate should plateau as a function of network complexity in order to tolerate noise in support of lossy compression. \rev{To evaluate this prediction, we operationalized network complexity supporting low fidelity communication by using path transitivity. Path transitivity quantifies the number of indirect pathways along the shortest path which are a one-connection longer detour (\textbf{Figure \ref{fig6}A}, Method \ref{pathTransitivity}). Along the shortest paths, the transitivity is the number of triangles formed by one edge in the shortest path and two connected edges composing an indirect pathway which exits and immediately returns to the shortest path. Operationalizing low fidelity with path transitivity stemmed from interpreting path transitivity using our model assumptions of random walk dynamics. }\\
	
	\rev{Applying our random walk model assumptions to path transitivity, if the shortest path represents the structure supporting highest fidelity because longer paths introduce information loss, then path transitivity's indirect pathways which are just one connection longer than the shortest path are the next-best paths for fidelity. Thus, path transitivity quantifies indirect pathways which offer a random walker the best \textit{approximations} of the highest fidelity path, or the best lossy compression. To operationalize} the complexity of the communication system contributing to low-fidelity transmission (better able to tolerate noise in support of lossy compression), we measured the number of nodes in the indirect pathways of path transitivity, which we termed shortest path complexity. \rev{In testing the prediction that information rate should plateau as a function of network complexity in low-fidelity regimes, we found} that the number of \rev{random walkers} plateaued non-linearly as a function of shortest path complexity, consistent with low-fidelity communication (\textbf{Figure \ref{fig6}C}). Model selection criteria support the non-linear form compared to a linear version of the same model (non-linear $AIC=7902$, linear $AIC=7915$; non-linear $BIC=7964$, linear $BIC=7968$). The non-linear fit of these data suggest that path transitivity supports low-fidelity communication that is tolerant to noise. 
	
	\subsubsection*{Hierarchical organization flexibly and efficiently supports both a high-fidelity and a low-fidelity regime}
	
	\rev{Our findings suggest that high fidelity depends on network \textit{size} while low  fidelity depends on network \textit{transitivity}. Prior work has shown that hierarchical organization supports the simultaneous presence of networks of \textit{large size and high \textit{transitivity}} \cite{ravasz2003hierarchical}. Therefore, we hypothesized that if the brain network has hierarchical organization, then such organization may enable flexible switching between high- and low-fidelity regimes. Hierarchical organization, wherein submodules are nested into successively larger but less densely interconnected modules, is thought to support efficient spatial embedding as well as specialized information transfer (\textbf{Figure \ref{fig6}D}, left) \cite{bassett2010efficient}. We first sought to assess if brain networks exhibit hierarchical organization. Hierarchical organization imposes a strict scaling law between transitivity and degree; the slope of this relationship can be used to identify the presence of hierarchically modular organization in real networks. This strict scaling is distinctive of hierarchical networks because increasing the size of a generic non-hierarchical network containing highly connected hubs (degree) will tend to diminish transitivity (clustering); however, hierarchical organization is known to decouple the size and transitivity of a network, which allows each property to vary individually \cite{ravasz2003hierarchical}. The characteristic scaling of reduced transitivity of brain regions with higher degree was present across the 5 brain network sizes (83 nodes: $t=-15.31$, $\textrm{bootstrap 95\% CI}$ $[-19.37, -12.29]$, $p<0.001$, $R^2=0.74$; 129 nodes: $t=-14.82, \textrm{bootstrap 95\% CI}$ $[-18.48, -11.93]$, $p<0.001, R^2=0.63$; 234 nodes: $t=-16.96$, $\textrm{bootstrap 95\% CI}$ $[-20.43, -13.75], p<0.001, R^2=0.55$; 463 nodes: $t=-26.47$, $\textrm{bootstrap 95\% CI}$ $[-30.02, -23.16]$, $p<0.001, R^2=0.60$; 1015 nodes: $t=-41.88,$ bootstrap 95\% CI $[-46.10, -38.11], p<0.001, R^2=0.63$; $F=917.3, df=1913, p<0.001$; \textbf{Figure \ref{fig6}D}, right). Hence, the connectome exhibits hierarchical organization.}\\ 
	
	\rev{We next assessed the hypothesis that hierarchical organization permits networks to have simultaneously large size and transitivity: two core network properties hypothesized to underlie high- and low-fidelity communication regimes \cite{ravasz2003hierarchical}. Hence, we sought to test whether hierarchical organization exhibits the hallmarks of both high- and low-fidelity regimes. If hierarchical organization supports a high-fidelity regime, then the information rate will monotonically increase with the complexity of the communication system. In testing this hypothesis, we found that the scaling characteristic of hierarchical organization in brain networks was associated with monotonically greater information rate ($F=2002$, \textrm{bootstrap 95\% CI} [1946.6, 2061.4], $df=13389, p<0.001,  R^2=0.62, \textrm{bootstrap 95\% CI} [0.62, 0.63]$, \textbf{Figure \ref{fig6}E}). Considering the slope of the increasing rate, networks with greater hierarchical organization supported high fidelity communication more efficiently ($t=32.4$, $\textrm{bootstrap 95\% CI}$ $[29.96, 34.79]$, $p<0.001$) than larger networks ($t=2640.32, \textrm{bootstrap 95\% CI}$ $[2556.3, 2725.07]$,  $p<0.001$). Taken together, our findings suggest that hierarchical organization, which is already known to be spatially efficient \cite{bassett2010efficient}, also supports high-fidelity network communication and does so more efficiently than large networks without hierarchical organization.}\\
  
    \rev{To enable flexible switching, hierarchical organization should also support a low-fidelity regime. If path transitivity supports a low-fidelity regime better able to tolerate noise, then hierarchical organization may support a low-fidelity regime by allowing for greater global transitivity in hierarchical networks compared to non-hierarchical networks. In testing this prediction, we found that the hierarchical organization of the connectome exhibited greater transitivity than non-hierarchical scale-free networks (generated with Method \ref{networkNulls}); and this increased transitivity became more pronounced with increasing network size (size-by-hierarchical network type interaction $t=-37.22$, $\textrm{bootstrap 95\% CI}$ $[-40.57, -34.12]$, $p<0.001$; size-by-scale-free network type interaction $t=-125.32, \textrm{bootstrap 95\% CI}$ $[-131.71, -118.70]$, $p<0.001$; \textbf{Figure \ref{fig6}F}). Thus, hierarchical organization could contribute to the flexibility of efficient coding by preserving high network transitivity for low fidelity communication that prioritizes noise tolerance and large network size for high fidelity communication that prioritizes accuracy.}
	
	\subsubsection*{Brain regions disproportionately scale in relation to regional prioritization of network communication fidelity or lossy compression}
	
    \rev{Having found that hierarchical organization preserves high transitivity despite large network size, we sought to understand how transitivity is associated with the areal expansion of brain regions.} Evolutionarily new connections may support higher-order and flexible information processing, emerging from \rev{disproportionate expansion of the association cortex \cite{buckner2013evolution}. In contrast, brain regions that are disproportionately out-scaled by total brain expansion may save material, space, and metabolic resources. To explore how sender and receiver compression efficiency relates to cortical areal expansion, we used published maps of areal scaling; here, allometric scaling coefficients were defined by the non-linear ratios of surface area change to total brain size change over development. \rev{Brain regions that disproportionately expanded in relation to total brain size during} neurodevelopment tended to have greater transitivity ($r=0.26$, $df=358$, $p_{\textrm{SPIN, Holm-Bonferroni}}=0.001$), which may support noise-tolerant lossy compression with minimal loss of fidelity in association cortex regions thought to underpin higher-order and flexible information processing. Disproportionately expanding brain regions tended to have greater sender ($r=0.14$, $df=358$, $p_{\textrm{SPIN, Holm-Bonferroni}}=0.045$; \textbf{Figure \ref{fig6}H}) and reduced receiver ($r=-0.20$, $df=358$, $p_{\textrm{SPIN, Holm-Bonferroni}}=0.03$; \textbf{Figure \ref{fig6}H}) compression efficiency. While brain regions that disproportionately expanded tend to prioritize messages received with high fidelity, brain regions that are disproportionately out-scaled by total brain expansion tend to receive compressed messages.}

	\subsection*{\rev{Hypothesis 5: Integrative hubs}}
	
	The fifth and final hypothesis of our model posits that the structural hubs of the brain's highly interconnected rich club supports information integration of \rev{randomly walking} signals \cite{van2012high}. To explain the hypothesized information integration roles of rich-club structural hubs, we investigated the compression efficiency of messages \rev{randomly walking} into and out of hub regions compared to that of other regions. In order to identify the rich-club hubs, we computed the normalized rich-club coefficient and identified 43 highly interconnected structural hubs (\textbf{Figure \ref{fig8}A}). Next, we computed the send and receive compression efficiency of rich-club hubs compared to all other regions. In support of their hypothesized function, we found that the rich-club hubs required receiving fewer random walkers compared to other regions (Wilcox rank sum test, $W=12829, p<0.001$), suggesting prioritization of information compression (or integration). For the rich-club hub to transmit outgoing messages with a fidelity that is equivalent to the incoming messages, the rich-club hubs required sending more random walkers compared to other brain regions ($W=64, p<0.001$),

	\begin{figure}[H]
	\centering
	\includegraphics[width=1\columnwidth]{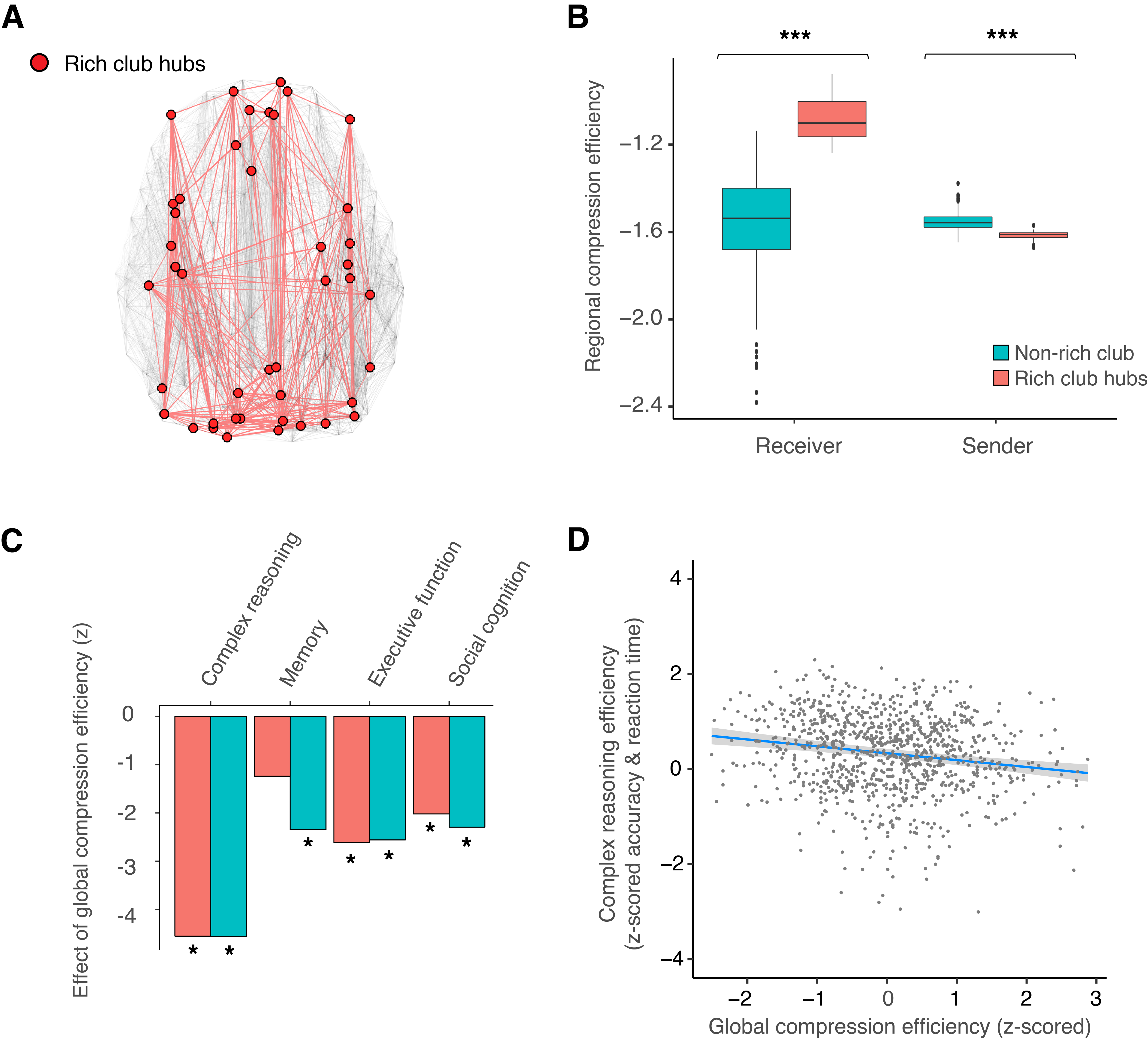}
		\caption[short]{\textbf{Compression efficiency explains the integrative role of rich-club hubs and individual differences in cognitive efficiency.} \textbf{\emph{(A)}} The rich club consists of highly interconnected structural hubs, thought to be the backbone of the brain connectome. Across participants, the rich club was comprised of 43 highly connected brain regions. \textbf{\emph{(B)}} The receiver compression efficiency of rich clubs was greater than other brain regions ($W=12829, \rev{\textrm{bootstrap 95\% CI} [9415.6, 16635.34],} p<0.001$), suggesting that the organization of the structural network prioritizes integration and compression of information arriving to rich-club hubs. The sender compression efficiency of rich clubs was reduced compared to other brain regions, supporting the fidelity of information broadcasting ($W=64, \rev{\textrm{bootstrap 95\% CI} [13.0, 138.03]}, p<0.001$). \textbf{\emph{(C)}} Individuals having non-rich-club regions with decreased compression efficiency tended to exhibit greater efficiency (\rev{combined speed and accuracy}) of complex reasoning (all $p$-values corrected using the Holm-Bonferroni method; $t=-4.72$, \rev{bootstrap 95\% CI [-6.63, -2.72]}, \rev{model adjusted $R^2$=0.21, bootstrap 95\% CI [0.16, 0.26]}, estimated model $df=10.50$, $p=2\times10^{-5}$), executive function ($t=-2.85$, \rev{bootstrap 95\% CI [-4.85, -1.0]}, \rev{model adjusted $R^2$=0.50, bootstrap 95\% CI [0.46, 0.56]}, estimated model $df=10.64$, $p=0.03$), and social cognition ($t=-2.30$, \rev{bootstrap 95\% CI [-4.21, -0.33]}, \rev{model adjusted $R^2$=0.22, bootstrap 95\% CI [0.17, 0.26]}, estimated model $df=10.45$, $p=0.04$), controlling for age, sex, age-by-sex interaction, degree, density, and in-scanner motion. Non-rich-club compression efficiency was not associated with memory efficiency ($t=-1.61$, \rev{bootstrap 95\% CI [-3.29, 0.07]}, estimated model $df=9.85$, $p=0.11$). Individuals with less rich-club compression efficiency tended to exhibit greater efficiency of complex reasoning (\rev{$t=-4.73$, bootstrap 95\% CI [-6.58, -2.82]}, \rev{model adjusted $R^2$=0.21, bootstrap 95\% CI [0.16, 0.26]}, $p=1\times10^{-7}$), memory efficiency (\rev{$t=-2.60$, bootstrap 95\% CI [-4.41, -0.79]}, \rev{model adjusted $R^2$=0.24, bootstrap 95\% CI [0.16, 0.26]}, $p=0.03$), social cognition }
		\label{fig8}
	\end{figure}
	
	\begin{figure}[H]
		\ContinuedFloat
        \caption[]{\textbf{continued}. (\rev{$t=-2.55$, bootstrap 95\% CI [-4.50, -0.73]}, \rev{model adjusted $R^2$=0.22, bootstrap 95\% CI [0.16, 0.25]}, $p=0.02$), and executive function (\rev{$t=-2.80$, bootstrap 95\% CI [-4.75, -0.87]}, \rev{model adjusted $R^2$=0.50, bootstrap 95\% CI [0.45, 0.56]}, $p=0.03$). Overall, individuals with brain structural networks prioritizing fidelity tended to perform with greater accuracy and/or speed in various cognitive functions. Asterisks depict significance following family-wise error correction. \textbf{\emph{(D)}} Individual differences in complex reasoning efficiency were negatively associated with individual differences in compression efficiency ($t = -4.95$, estimated model $df=11.52$, $p < 0.001$; confidence intervals displayed as the gray band), controlling for global efficiency, age, sex, age-by-sex interaction, degree, density, and in-scanner motion. See \textbf{Supplementary Figure} \textbf{9} for additional scatterplots of the correlation between cognitive efficiency and compression efficiency, as well as for separated metrics of accuracy and speed. For comparison to a commonly used metric of shortest-path information integration, we include global efficiency in the model, which was positively correlated with complex reasoning efficiency ($t = 2.68$, estimated model $df=11.52$, $p < 0.01$). Global efficiency was not partially correlated with compression efficiency ($r=-0.04, df=997, p=0.18$), controlling for age, sex, age-by-sex interaction, degree, density, and motion. Hence, reduced compression efficiency prioritizing fidelity explains individual differences in complex reasoning efficiency.}
	\end{figure}
	
	supporting the notion that rich-club hubs serve as high-fidelity information broadcasting sources. The function of hubs apparently compensates for high metabolic cost \cite{collin2013structural}. However, we did not find evidence of greater CBF in rich-club hubs compared to non-hubs, suggesting that hubs may be more metabolically efficient than previously known (\textbf{Supplementary Figure 8}). This may potentially be due to hubs being compression efficient receivers (see \textbf{Supplementary Results} and \textbf{Discussion} for additional explanations). Though we did not find evidence of high metabolic cost, the contrasting roles of prioritizing input compression and output fidelity within rich-club hubs was consistent with our hypothesis and the current understanding of rich-club hubs as the information integration centers and broadcasters of the brain's network \cite{van2012high}.
	
	\subsubsection*{Compression efficiency predicts behavioral performance}
	
	\rev{Efficient coding predicts that the brain should not only balance fidelity and information compression as quantified by compression efficiency, but should do so in a way that improves future behavior \cite{niven2007fly, chalk2018toward}. Thus, we next sought to evaluate the association between compression efficiency in rich-club hubs and cognitive performance in a diverse battery of tasks. In light of trade-offs between communication fidelity} \rev{and information compression, we hypothesized that compression efficiency would correlate with cognitive efficiency, defined as the combined speed and accuracy of task performance. Compression efficiency prioritizing lossy compression and low-dimensionality should predict worse performance \cite{rigotti2013importance}. To assess the relationships between compression efficiency and cognitive efficiency, we used four independent cognitive domains that have been established by confirmatory factor analysis to assess individual variation in tasks of complex reasoning, memory, executive function, and social cognition (Method \ref{cognitiveBattery}). Consistent with the hypothesis, we found that individuals having rich-club hubs with reduced compression efficiency, prioritizing transmission fidelity, tended to exhibit increased cognitive efficiency of complex reasoning (all $p$-values corrected using the Holm-Bonferroni family-wise error method; $t=-4.73$, bootstrap 95\% CI [-6.58, -2.82], model adjusted $R^2$=0.21, bootstrap 95\% CI [0.16, 0.26], estimated model $df=10.59$, $p=2\times10^{-7}$), memory ($t=-2.60$, bootstrap 95\% CI [-4.41, -0.79], model adjusted $R^2$=0.24, bootstrap 95\% CI [0.16, 0.26], estimated model $df=9.85$, $p=0.03$), executive function ($t=-2.80$, bootstrap 95\% CI [-4.75, -0.87], model adjusted $R^2$=0.50, bootstrap 95\% CI [0.45, 0.56], estimated model $df=10.69$, $p=0.03$), and social cognition ($t=-2.55$, bootstrap 95\% CI [-4.50, -0.73], model adjusted $R^2$=0.22, bootstrap 95\% CI [0.16, 0.25], estimated model $df=10.47$, $p=0.02$). See \textbf{Supplementary Figure 9} for similar non-hub correlations, as well as speed and accuracy modeled separately. Our finding that the compression efficiency of rich-club hubs was associated with cognitive efficiency is consistent with the notion that the integration and broadcasting function of hubs contributes to cognitive performance \cite{van2012high}. Importantly, compression efficiency explained variation in cognitive efficiency even when controlling for the commonly used shortest-path measure of global efficiency (\textbf{Figure \ref{fig8}D}; compression efficiency $t = -4.95$, estimated model $df=11.52$, $p < 0.001$; global efficiency $t = 2.68$, estimated model $df=11.52$, $p < 0.01$). \rev{Collectively, individuals} with connectomes prioritizing fidelity tended to perform with greater cognitive efficiency in a diverse range of functions, consistent with the understanding of high-dimensionality neural representations predicting flexible behavioral performance \cite{van2012high, rigotti2013importance}.}
		
	\section*{Discussion}
	
	\rev{The efficient coding principle and rate-distortion theory constrain models of brain network communication, explaining how connectome biology and architecture prioritize either communication fidelity or lossy compression \cite{levy1996energy, laughlin2001energy, achard2007efficiency, johansen2010behavioural, bullmore2012economy, goni2013exploring, palmer2015predictive, avena2015network, rubinov2016constraints, avena2017path, avena2018communication}. Information compression can improve prediction and generalization, separate relevant features of information, and efficiently utilize limited capacity by balancing the unreliability of stochasticity with the redundancy in messages \cite{10.2307/j.ctt17kk982, olshausen2004sparse, palmer2015predictive, sims2016rate, sims2018efficient}. \revtwo{We observed a rate-distortion gradient for every individual, consistent with the conceptualization of the brain structural network as a communication channel with limited capacity. While each individual's network adheres to rate-distortion theory, differences in structural connectivity lead to substantial variance in the compression efficiency. In addition to some of the variation being explained by age and sex, our results also indicate that variance could be driven by biological differences, such as in metabolic expenditure and myelin, as well as topological differences, such as transitivity, degree, and hierarchical organization.} We found that biological investments of metabolic resources and myelin in connectome topology, hierarchically interconnected brain regions, and highly connected network hubs each supported efficient coding and compressed communication dynamics. We found evidence consistent with five predictions of rate-distortion theory adapted from prior literature, corroborating the validity of \rev{this theoretical model of macroscale efficient coding} \cite{van2012high, marzen2017evolution, sims2018efficient}.} ~\\
	 
    In addition to these findings, we reported two important null results which could be explained in part by the macroscale efficient coding model. First, we did not find evidence for the hypothesis that metabolic savings are associated with shortest path routing communication dynamics, a key redeeming feature against acknowledged theoretical shortcomings of this routing model \cite{varkuti2011quantifying, bullmore2012economy, avena2018communication}. Rather, we found evidence of efficient coding, implemented using random walks, as in other areas of neuroscience \cite{barlow1961possible, laughlin1998metabolic, laughlin2001energy, 10.2307/j.ctt17kk982, olshausen2004sparse, wei2015bayesian, palmer2015predictive, deneve2017brain, chalk2018toward, weber2019coding}. Efficient coding generated unique predictions of communication as a stochastic process governed by trade-offs between lossy compression and transmission fidelity \cite{mackay2003information}. The evidence we presented was consistent with these predictions and would be difficult to explain with alternative models of shortest-path routing (see \textbf{Supplementary Modeling/Math Notes} and \textbf{Discussion}). This null result challenges the validity of the shortest path routing model as anchoring one end of a hypothesized spectrum of communication mechanisms and the usage of shortest path routing metrics to quantify the integrative capacity of particular brain areas or whole brain connectivity \cite{avena2018communication}. Rather, the macroscale efficient coding model implemented by random walk dynamics provides strong theoretical interpretation of information integration in hubs as lossy compression \cite{van2012high, chanes2016redefining}. \\
	
	Second, we did not find evidence for high metabolic cost associated with structural network hubs. Rather, we found that hubs are regions that receive inputs with greater compression efficiency; regions prioritizing input compression tend to be disproportionately out-scaled by total brain expansion during development, consistent with cortical consolidation (thinning and myelination) \cite{whitaker2016adolescence}. In contrast, regions prioritizing input fidelity tend to disproportionately expand in relation to total brain growth during development, consistent with theories positing that evolutionary expansion and new connections support the flexibility of neural activity for higher-order cognition \cite{goldman1988topography, bullmore2012economy, buckner2013evolution,glasser2014trends, scholtens2014linking, reardon2018normative, vij2018evolution, Theodoni2020.02.28.969824}. Structural network hubs support unique information integrative processes, apparently offsetting high metabolic, spatial, and material costs \cite{van2012high, vertes2014generative, crossley2014hubs, scholtens2014linking, chanes2016redefining, oldham2018development, 10.1371/journal.pcbi.1006833}. Hence, hub dysfunction may be particularly costly. Our null result motivates reconsideration of metabolic costs, at least for the simplest and most common definition of a structural connectivity hub using the measure of degree centrality (see \textbf{Supplementary Figure 8}, \textbf{Results}, and \textbf{Discussion}) \cite{oldham2018development}. \rev{An alternative hypothesis to high metabolic cost of hubs is that long-run metabolic savings depend on frequent usage of hubs to compensate for the expense of maintaining their size and connectivity \cite{wang2008functional, harris2012energetics}.} Although further investigation of the metabolic costs of rich-club hubs is warranted, our findings nevertheless reinforce a wealth of evidence emphasizing the importance of the development, resilience, and function of hubs in cognition and psychopathology \cite{van2012high,crossley2014hubs, liang2013coupling, mivsic2015cooperative, chanes2016redefining, whitaker2016adolescence, scholtens2014linking, gollo2018fragility}. ~\\
	
	Our work admits several theoretical and methodological limitations. First, regionally aggregated brain signals are not discrete Markovian messages and do not have goals like reaching specific targets. As in recent work, our model is a deliberately simplified but useful abstraction of macroscale brain network communication \cite{mivsic2015cooperative}. Second, although we modeled random walk dynamics in light of prior methodological decisions and information theory benchmarks \cite{goni2013exploring}, compression efficiency can be implemented using alternative approaches. Several methodological limitations should also be considered. The accurate reconstruction of white-matter pathways using \rev{diffusion imaging} and tractography remains limited \cite{zalesky2016connectome}. \revtwo{While we chose to model structural connections using fractional anisotropy because it is one of the most widely used measures of white matter microstructure, future research could evaluate these same hypotheses in structural networks built from different measures and of other species \cite{jones2013white, Chang2017TheRO, teich2021crystallinity}.} Moreover, non-invasive measurements of CBF with high sensitivity and spatial resolution remain challenging. However, we acquired images using an ASL sequence providing greater sensitivity and approximately four times higher spatial resolution than prior developmental studies of CBF \cite{satterthwaite2014impact}. Lastly, our data was cross-sectional, limiting the inferences that we could draw about neurodevelopmental processes. ~\\
	
	In summary, our study advances understanding of how brain metabolism and architecture at the macroscale supports communication dynamics in complex brain networks as efficient \rev{coding}. Our model provides a simple framework for investigating efficient coding at the whole brain network level and an important basis to build towards computational network models of special cases of efficient coding, such as robust, sparse, and predictive coding \cite{chalk2018toward}. \rev{In quantifying integrative communication dynamics and lossy compression of the macroscale brain network, these results are relevant to future research on low-dimensional neural representations \cite{rigotti2013importance, shine2019human, tang2019effective}, efficient control of functional network dynamics \cite{tang2017developmental, srivastava2020models}, and hierarchical abstraction of behaviorally relevant cortical representations \cite{schapiro2017complementary, stachenfeld2017hippocampus, momennejad2020learning}.} Future research could investigate whether other biological properties of the network support efficient coding, such as gradients in brain structure and function \cite{barbas1986pattern, kingsbury2001cortex, charvet2014evo, charvet2015systematic, huntenburg2018large, paquola2020cortical, vazquez2020signal}. \revtwo{Neurodevelopmental processes vary with the naturalistic environment and socioeconomic background, suggesting that they are shaped by exposure to different experiences and expectations \cite{tooley2021environmental}. We hypothesize that the neurodevelopmental range of compression efficiency reflects a relationship between source coding (statistics or compressibility of the naturalistic input information) and channel coding (compression efficiency), which demarcate zones of possible brain network communication \cite{olshausen2004sparse, mackay2003information}.} Lastly, the compression efficiency metric offers a novel tool to test leading hypotheses of dysconnectivity \cite{di2014unraveling}, hubopathy \cite{crossley2014hubs, gollo2018fragility}, disrupted information integration \cite{hernandez2015neural, chanes2016redefining}, and neural noise \cite{dinstein2012unreliable} in neuropsychiatric disorders.
	
	\section*{Author Contributions}
	
	D.Z wrote the paper. C.W.L., T.D.S., and D.S.B. edited the paper. D.Z. developed the theory with input from C.W.L, T.D.S., and D.S.B. R.C., G.L.B., and Z.C. preprocessed the data. D.Z. performed the analysis with input from Z.C. T.M.M. performed data preprocessing and interpretation of statistical models. D.R. preprocessed the data. J.D. developed imaging acquisition methods. R.E.G. acquired funding for data collection and performed data collection. R.C.G. provided expertise in cognitive phenotyping. D.Z., T.D.S, and D.S.B. designed the study. D.S.B. and T.D.S. acquired funding to support theory development and data analysis, and contributed to theory and data interpretation.
		
	\section*{Acknowledgments}
	
	We acknowledge helpful discussions with Dr. Jennifer Stiso, Dr. Richard Betzel, Dr. David Lydon-Staley, Dr. Lorenzo Caciagli, Adon Rosen, and Dr. Bart Larsen. The work was largely supported by the John D. and Catherine T. MacArthur Foundation, the ISI Foundation, the Paul G. Allen Family Foundation, the Alfred P. Sloan Foundation, the NSF CAREER award PHY-1554488, NIH R01MH113550, NIH R01MH112847, and NIH R21MH106799. Secondary support was also provided by the Army Research Office (Bassett-W911NF-14-1-0679, Grafton-W911NF-16-1-0474) and the Army Research Laboratory (W911NF-10-2-0022). D.Z. acknowledges support from the National Institute of Mental Health F31MH126569. C.W.L. acknowledges support from the James S. McDonnell Foundation 21$^{\text{st}}$ Century Science Initiative Understanding Dynamic and Multi-scale Systems - Postdoctoral Fellowship Award. The content is solely the responsibility of the authors and does not necessarily represent the official views of any of the funding agencies. 
	
	\section*{Competing Interests}
	
	The authors declare that they have no competing interests.
	
	\section{Materials and Methods}
	\subsection{Participants}
	
	As described in detail elsewhere \cite{satterthwaite2014neuroimaging}, diffusion \rev{weighted} imaging (DWI) and arterial-spin labeling (ASL) data were acquired for the Philadelphia Neurodevelopmental Cohort (PNC), a large community-based study of neurodevelopment. The subjects used in this paper are a subset of the 1,601 subjects who completed the cross-sectional imaging protocol. We excluded participants with health-related exclusionary criteria (n=154) and with scans that failed a rigorous quality assurance protocol for DWI (n=162) \cite{roalf2016impact}. We further excluded subjects with incomplete or poor ASL and field map scans (n=60). Finally, participants with poor quality T1-weighted anatomical reconstructions (n=10) were removed from the sample. The final sample contained 1042 subjects (mean age=15.35, SD=3.38 years; 467 males, 575 females). Study procedures were approved by the Institutional Review Board of the Children's Hospital of Philadelphia and the University of Pennsylvania. All adult participants provided informed consent; all minors provided assent and their parent or guardian provided informed consent. ~\\
	
	\subsection{Cognitive Assessment} \label{cognitiveBattery}
	
	All participants were asked to complete the Penn Computerized Neurocognitive Battery (CNB). The battery consists of 14 tests adapted from tasks typically applied in functional neuroimaging, and which measure cognitive performance in four broad domains \cite{satterthwaite2014neuroimaging}. The domains included: (1) executive control (i.e., abstraction and flexibility, attention, and working memory), (2) episodic memory (i.e., verbal, facial, and spatial), (3) complex cognition (i.e., verbal reasoning, nonverbal reasoning, and spatial processing), (4) social cognition (i.e., emotion identification, emotion intensity differentiation, and age differentiation), and (5) sensorimotor and motor speed. Performance was operationalized as $z$-transformed accuracy and speed. The speed scores were multiplied by $-1$ so that higher indicates faster performance, and efficiency scores were calculated as the mean of these accuracy and speed $z$-scores. The efficiency scores were then $z$-transformed again, to achieve $\mathrm{mean}=0$ and $\mathrm{SD}=1.0$ for all scores. Confirmatory factor analysis supported a model of four latent factors corresponding to the cognitive efficiency of executive function, episodic memory, complex cognition, and social cognition \cite{moore2015psychometric}. Hence, we used these four cognitive efficiency factors in our analyses. \rev{In a factor solution separately modeling accuracy and speed, the accuracy factors correspond to (1) executive and complex cognition, (2) social cognition, and (3) memory. The speed factors correspond to (1) fast speed (e.g., working memory and attention tasks requiring constant vigilance), (2) episodic memory speed, and (3) slow speed (e.g., tasks requiring complex reasoning).} 
	
	\subsection{Image Acquisition, Preprocessing, and Network Construction}
	
	Neuroimaging acquisition and pre-processing were as previously described \cite{satterthwaite2014neuroimaging}. We depict the overall workflow of the neuroimaging and network extraction pipeline in \textbf{\rev{Supplementary} Figure 1}.
	
	\subsubsection{Diffusion Weighted Imaging}
	
	As was previously described \cite{baum2017modular, tang2017developmental}, \rev{diffusion imaging} data and all other MRI data were acquired on the same 3T Siemens Tim Trio whole-body scanner and 32-channel head coil at the Hospital of the University of Pennsylvania. D\rev{W}I scans were obtained using a twice-focused spin-echo (TRSE) single-shot EPI sequence (TR = 8100 ms, TE = 82 ms, FOV = 240 mm$^2$/240 mm$^2$; Matrix = RL: 128/AP:128/Slices:70, in-plane resolution (x \& y) 1.875 mm$^2$; slice thickness = 2 mm, gap = 0; FlipAngle = 90°/180°/180°, volumes = 71, GRAPPA factor = 3, bandwidth = 2170 Hz/pixel, PE direction = AP). The sequence employs a four-lobed diffusion encoding gradient scheme combined with a 90-180-180 spin-echo sequence designed to minimize eddy current artifacts. The complete sequence consisted of 64 diffusion-weighted directions with $b$ = 1000 s/mm$^2$ and 7 interspersed scans where $b$ = 0 s/mm$^2$. Scan time was about 11 min. The imaging volume was prescribed in axial orientation covering the entire cerebrum with the topmost slice just superior to the apex of the brain \cite{roalf2016impact}.
	
	\subsubsection{Connectome construction}
	
	Cortical gray matter was parcellated according to the Glasser atlas \cite{glasser2016multi}, defining 360 brain regions as nodes for each subject's structural brain network, denoted as the weighted adjacency matrix $\mathbf{A}$. To assess multiple spatial scales, cortical and subcortical gray matter was parcellated according to the Lausanne atlas \cite{cammoun2012mapping}. Together, 89, 129, 234, 463, and 1015  dilated brain regions defined the nodes for each subject’s structural brain network in the analyses of \textbf{Figure \ref{fig6}}. ~\\

	D\rev{W}I data was imported into DSI Studio software and the diffusion tensor was estimated at each voxel \cite{yeh2013deterministic}. For deterministic tractography, whole-brain fiber tracking was implemented for each subject in DSI Studio using a modified fiber assessment by continuous tracking (FACT) algorithm with Euler interpolation, initiating 1,000,000 streamlines after removing all streamlines with length less than 10mm or greater than 400mm. Fiber tracking was performed with an angular threshold of 45, a step size of 0.9375mm, and a fractional anisotropy (FA) threshold determined empirically by Otzu’s method, which optimizes the contrast between foreground and background \cite{yeh2013deterministic}. FA was calculated along the path of each reconstructed streamline. For each subject, edges of the structural network were defined where at least one streamline connected a pair of nodes. Edge weights were defined by the average FA along streamlines connecting any pair of nodes. \rev{The resulting structural connectivity matrices were not thresholded and contain edges weighted between 0 and 1.}  ~\\

	\subsubsection{Arterial-Spin Labeling}
	
	CBF was quantified from control-label pairs using ASLtbx \cite{wang2008empirical}, as was previously described \cite{satterthwaite2014impact}. We consider $f$ as CBF, $\delta M$ as the difference of the signal between the control and label acquisitions, $R_{1a}$ as the longitudinal relaxation rate of blood, $\tau$ as the labeling time, $\omega$ as the post-labeling delay time, $\alpha$ as the labeling efficiency, $\lambda$ as the blood/tissue water partition coefficient, and $M_0$ as the approximated control image intensity. Together, CBF $f$ can be calculated according to the equation:
	
	\begin{equation}
	f=\frac{\Delta M \lambda R_{1 a} \exp \left(\omega R_{1 a}\right)}{2 M_{0} \alpha}\left[1-\exp \left(-\tau R_{1 a}\right)\right]^{-1}. 
	\end{equation}
	
	Because prior work has shown that the T1 relaxation time changes substantially in development and varies by sex, this parameter was set according to previously established methods, which enhance CBF estimation accuracy and reliability in pediatric populations \cite{wu2010vivo, jain2012longitudinal}. \rev{As in prior work, the global network CBF was calculated as the average CBF across all brain regions to obtain an individual participant's CBF \cite{satterthwaite2014impact}.}
	
	\subsection{Brain Maps}
	
	\subsubsection{Cortical Myelin}
	
	As described previously \cite{glasser2011mapping}, cortical myelin content was calculated by dividing the T1w image signal by the T2w image signal. Specifically, we define the myelin content $x^2$ in the following manner:
	
	\begin{equation}
	\frac{\mathrm{T} 1 \mathrm{w}}{\mathrm{T} 2 \mathrm{w}} \approx \frac{x * b}{(1 / x) * b}=x^{2},
	\end{equation}
\noindent where $x$ is the myelin contrast in the T1w image, $1/x$ is the myelin contrast in the T2w, and $b$ is the receive bias field in both T1w and T2w images. We used a published atlas generated by this method \cite{glasser2014trends}.

	\subsubsection{Cortical Areal Scaling}
	
	As described previously \cite{reardon2018normative}, to estimate cortical areal scaling between the size of cortical regions and the total brain, regression coefficients $\beta$ were estimated for log$_{10}$(total cortical surface area) as a covariate predicting log$_{10}$(vertex area) using spline regression models that incorporated effects of age and sex on vertex area \cite{wood2004stable}. We used the following relational form:
	
	\begin{equation}
	\log 10(\text {Vertex area}) \sim \mathrm{s}(\text {age by }=\mathrm{sex})+\mathrm{B}_{\mathrm{I}}\left[\log 10\left(\text{total}_{-} \text {area}\right)\right].
	\end{equation}
	
When $\beta$ is 1, the scaling between total brain size and brain regions is linear. When $\beta$ deviates greater or less than 1, scaling is non-linearly and disproportionately expanding or contracting. We used the published atlas generated using the same data as in our study \cite{reardon2018normative, satterthwaite2014neuroimaging}.
	
	\subsection{Network Statistics}
	
	\subsubsection{Global Efficiency}
	
	In the context of the brain structural connectome, global efficiency represents the strength of the shortest paths between brain regions supporting efficient communication. In network neuroscience, global efficiency is commonly used as a metric of a brain network's capacity for shortest path routing \cite{bullmore2012economy,goni2013exploring,avena2018communication}. We calculated the common global efficiency statistic \cite{latora2001efficient}, which is defined for a graph $G$ as:
	\begin{equation}\label{globEffEq}
	\mathcal{E}_{glob}(G) = \frac{1}{\mathcal{N}(\mathcal{N}-1)}\sum_{i \neq j \epsilon G}\frac{1}{d_{ij}},
	\end{equation}
	\noindent where $\mathcal{N}$ is the number of nodes and $d_{ij}$ is the shortest distance between node $i$ and node $j$. Intuitively, a high $\mathcal{E}$ value indicates greater potential capacity for global and parallel information exchange along shortest paths, and a low $\mathcal{E}$ value indicates decreased capacity for such information exchange \cite{latora2001efficient}.
	
	\subsubsection{Path Strengths} \label{pathStrengths}
	
	Beyond shortest paths between pairs of brain regions, we also sought to measure the strength of structural connections $S$ comprising the paths of multiple connections. As global efficiency measures the capacity of brain networks for shortest path routing, path strengths measure the capacity for \rev{stochastic communication}. Path strengths are apt for assessing the network capacity for \rev{a model of stochastic transmission of impulses} because paths can be represented as random walks $p=\left(i, j, \ldots, k\right)$, where $p$ is a path and $i$, $j$, and $k$ are nodes in the path. As in prior work \cite{becker2018spectral}, the strength of the weighted connections in a path, denoted $\omega(p)$, in the graph $G$ with adjacency matrix $\mathbf{A}$ is defined as:
	\begin{equation}
	\omega(p)=[\mathbf{A}]_{i_{0} i_{1}}[\mathbf{A}]_{i_{1} i_{2}} \ldots[\mathbf{A}]_{i_{(l-1)} i_{i}},
	\end{equation}
	\noindent where the matrix products produce the strengths of all possible random walks according to the length of $p$, as depicted in the schematic \textbf{Figure \ref{fig1}B}. Then, for walks of length $n$, the strengths of the paths from node $i$ to node $j$ are defined as:
	\begin{equation} \label{eqPathStrength}
	\left[\mathbf{A}^{n}\right]_{i j}=\sum_{p \in \mathcal{P}_{i j}^{n}} \omega(p),
	\end{equation}
	\noindent where ${P}_{i j}^{n}$ is the set of all walks from node $i$ to node $j$ with length $n$. When $n$=1, the matrix exponent produces a matrix with elements equal to $d_{ij}$ from Equation 1, or the shortest distance between node $i$ and node $j$. Intuitively, a high path strength represents structural paths that consist of higher integrity connections measured by D\rev{W}I, whereas a low path strength indicates paths consisting of low integrity connections. To compute node strengths, the values for each node were summed. An average value was also calculated across node strengths per individual participant.
	
	\subsubsection{Path Transitivity} \label{pathTransitivity}
	
	Shortest paths confer advantages in speed and signal fidelity when messages are transmitted by \rev{random walks}. Therefore, we sought to measure a property of brain network architecture supporting \rev{random walks} by shortest paths. Local detours which first leave and then re-access the shortest path serve to support such {random walks}, and the potential for such local detours can be estimated using a measure called path transitivity (see \textbf{Figure \ref{fig6}A}, left) \cite{goni2014resting}. Path transitivity was previously used to predict functional BOLD activation comparably to conventional distance or computational models of neural dynamics. To compute path transitivity, we first calculated the matching index for each pair of successive nodes $i$ and $j$ along the shortest path $\pi_{s \rightarrow t}$, with neighboring non-shortest path nodes $k$ as:
	\begin{equation} 
	m_{i j}=\frac{\sum_{k \neq i, j}\left(w_{i k}+w_{j k}\right) \Theta\left(w_{i k}\right) \Theta\left(w_{j k}\right)}{\sum_{k \neq j} w_{i k}+\sum_{k \neq i} w_{j k}},
	\end{equation}
	\noindent where $w$ is the connection weight, and $\Theta(w_{i k})=1$ if $w_{i k}>0$, and 0 otherwise. Intuitively, the numerator is non-zero if and only if there are two locally detouring connections that make a closed triangle along the shortest path. If either of the two connections $w_{i k}$ or $w_{j k}$ does not exist, then the numerator is 0. With the denominator representing the strength of all cumulative connections of the shortest path nodes, the matching index fraction then represents the density of closed triangles (i.e., transitivity) around the shortest path. \\
	
	Whereas the matching index is a pairwise measure of the density of locally returning detours, path transitivity generalizes the density across the shortest path. Using the computed matching index $m_{ij}$ for each pairwise connection $\Omega$ from source node $s$ to target node $t$ by the set of shortest path edges $\pi_{s \rightarrow t}$, we compute path transitivity $M$ as:
	\begin{equation}\label{eqPathTrans}
	M\left(\pi_{s \rightarrow t}\right)=\frac{2 \sum_{i \in \Omega} \sum_{j \in \Omega} m_{i j}}{|\Omega|(|\Omega|-1)},
	\end{equation}
	\noindent where the numerator sums the matching index $m_{ij}$ for all edges in $\Omega$, the scale factor of 2 indicates an undirected graph, and the denominator sums over all possible edges. Intuitively, a high path transitivity $M$ indicates that the shortest path is more densely encompassed by locally detouring triangular motifs. Low path transitivity indicates that the shortest path is surrounded by connections that deviate from the shortest path without an immediate avenue of return. \rev{An individual-level value of path transitivity was calculated as the average path transitivity across brain regions.}
	
	\subsubsection{Modularity}
	
	Modularity is a common architectural feature observed in neural systems across species. A single community contains brain regions that are more highly connected to each other than to brain regions located in other communities (see \textbf{Supplementary Figure 2A}, right). Modularity of brain networks is spatially efficient, supports the development of executive function in youths, and supports flexibly adaptable functional activations according to distinct task demands \cite{sporns2016modular,bassett2010efficient,baum2017modular,bassett2011dynamic,bertolero2015modular}. To assess modularity, we apply a common community detection technique known as modularity maximization \cite{newman2006modularity}, in which we used a Louvain-like locally greedy algorithm \cite{blondel2008fast} to maximize a modularity quality function for the adjacency matrix $A$. The modularity quality function is defined as:
	\begin{equation} \label{eqModularity}
	Q = \frac{1}{2\mu}\sum_{ij}\left(A_{ij}-\gamma P_{ij} \right) \delta(g_i,g_j),
	\end{equation}
	where $\mu = \frac{1}{2}\sum_{ij}A_{ij}$ denotes the total weight of $\mathbf{A}$, $A_{ij}$ encodes the weight of an \emph{edge} between node $i$ and node $j$ in the structural connectivity matrix, $\mathbf{P}$ represents the expected strength of connections according to a specified null model \cite{newman2006modularity}, $\gamma$ is a structural resolution parameter that determines the size of modules, and $\delta$ is the Kronecker function which is 1 if $g_i = g_j$ and zero otherwise. As in prior work, we set $\gamma$ to the default value of $1$ \cite{bassett2011dynamic}. Intuitively, a high $Q$ value indicates that the structural connectivity matrix contains communities, where nodes within a community are more densely connected to one another than expected under a null model. Modularity maximization is commonly used to detect community structure, and to quantitatively characterize that structure by assessing the strength and number of communities \cite{baum2017modular,bassett2011dynamic,newman2006modularity}. \rev{Regional contributions to the modularity quality function were used for analyses of brain regions.}
	
	\subsubsection{Resource Efficiency} \label{resourceEfficiency}
	
	A signal that randomly walks along the shortest path between brain regions confers advantages in speed, reliability, and fidelity \cite{bullmore2012economy,2016207,avena2018communication}. Following prior work, we sought to compute the number of random walkers beginning at node $i$ that were required for at least one to travel along the shortest path to another node $j$ with probability $\eta$ \cite{goni2013exploring, 2016207}. To begin, we consider the transition probability matrix by $\mathbf{U}$, defined as $\mathbf{U}=\mathbf{W} \mathbf{L}^{-1}$, where each entry $W_{i j}$ of $\mathbf{W}$ describes the weight of the directed edge from node $i$ to node $j$, and each entry $L_{ii}$ of the diagonal matrix $\mathbf{L}$ is the strength of each node $i$, defined as $\sum_i W_{i j}$. Intuitively, each entry $U_{i j}$ of $\mathbf{U}$ defines the probability of a random walker traveling from node $i$ to node $j$ in one step. Next, to compute the probability that a random walker travels from node $i$ to node $j$ along the shortest path, we define a new matrix $U'(i)$ that is equivalent to $\mathbf{U}$ but with the non-diagonal elements of row $i$ set to zero and $U_{ii}=1$ as an absorbant state. Then, the probability of randomly walking from $i$ to $j$ along the shortest path is given by:
	
	\begin{equation}
	1 - \sum_{n = 1}^N [U'(i)^H]_{in},
	\end{equation}
	
	\noindent where $H$ is the number of connections composing the shortest path from $i$ to $j$. Similarly, the probability $\eta$ of releasing $r$ random walkers at node $i$ and having at least one of them reach node $j$ along the shortest path is given by:
	
	\begin{equation}
	\eta = 1 - \left( \sum_{n = 1}^N [U'(i)^H]_{in} \right)^r.
	\end{equation}
	
	\noindent Setting the above probability to some set value $\eta$, we can then solve for the number of random walkers $r$ required to guarantee (with probability $\eta$) that at least one of them travels from $i$ to $j$ along the shortest path, denoted by:
	
	\begin{equation} \label{eq9}
	r_{i j}(\eta)=\frac{\log (1-\eta)}{\log \left(\sum_{n=1}^{N}\left[U'(i)^{H}\right]_{in}\right)}.
	\end{equation}
	
	\noindent The number of random walkers $r_{i j}$ has been referred to as resources in prior literature \cite{goni2013exploring}. In our analyses, we calculate resources $r_{i j}$ over a range of values of $\eta$ for each participant. Finally, to calculate the resource efficiency of each participant, the resource efficiency of an entire network is taken to be $1/(r_{i j}(\eta))$ averaged over all pairs of nodes $i$ and $j$. With the right stochastic matrix $U'_{i}$, the resource efficiency of brain regions as message senders is $1/(r_{i j}(\eta))$ averaged over $i$, while brain regions as message receivers is $1/(r_{j i}(\eta))$ averaged over $j$. 
	
	\subsubsection{Compression Efficiency} \label{compressionEfficiency}
	
	Rate-distortion theory formalizes the study of information transfer as passing signals (messages) through a capacity-limited information channel. A signal $x$ is encoded as $\hat{x}$ with a level of distortion $D$ that depends on the information rate $R$. The greater the rate, the less the distortion. The rate-distortion function $R(D)$ defines the minimum information rate required to transmit a signal corresponding to a level of signal distortion (see \textbf{Figure \ref{fig4}A}). Lossy compression arises from the choice of the distortion function $d(x, \hat{x})$, which implicitly determines the relevant and irrelevant features of a signal. With the true signal $x$ mapped to the compressed signal $\hat{x}$ described by $p(\hat{x} | x)$, the rate-distortion function is defined by minimizing the mutual information of the signal and compression over the expected distortion defined as $d(x, \hat{x})_{p(x, \hat{x})}=\sum_{x \in X} \sum_{\hat{x} \in \hat{X}} p(x, \hat{x}) d(x, \hat{x})$:
	
	\begin{equation}
	R(D) \equiv \min_{\substack{d(x, \hat{x})}} I(X,\hat{X}) =\sum_{x \in \Omega_{X}} \sum_{\hat{x} \in \Omega_{\hat{X}}} P(\hat{x} | x) P(x) \log _{2}\left(\frac{P(\hat{x} | x)}{P(\hat{x})}\right).
	\end{equation}
	
	\noindent By minimizing the mutual information $I(X,\hat{X})$, we arrive at a probabilistic map from the signal to the compressed representation, where the information gain between the signal and compression is as small as possible (i.e., high fidelity) to favor the most compact representations. \\
	
	Similar to the mathematical framework of rate-distortion theory, we sought to specify a distortion function reflecting communication over the brain's structural network. Prior work building models of perceptual and cognitive performance have inferred distortion functions through Bayesian inference of a loss function \cite{sims2016rate,sims2018efficient}. For instance, the loss function could be the squared error denoting the residual values of the true signal minus the compression, $\textit{L}=(\hat{x}-x)^2$ (\textbf{Figure \ref{fig4}A}). A neural rate-distortion theory has been theoretically developed \cite{marzen2017evolution}, but remains empirically untested due in part to a lack of methodological tools at the level of brain systems. Moreover, it has been difficult to define a distortion function that incorporates both true signals $x$ and compressed signals $\hat{x}$ in part because the measurements of these signals in human brain networks remains challenging. Here, we define an analogous framework of information transfer through capacity-limited channels in the structural network of the brain. Particularly, we build a distortion function from the simple intuition that the shortest path is the route that most reliably preserves signal fidelity, as depicted in (\textbf{Figure \ref{fig4}B}).  \\
	
	Given that a random walker propagating from node $i$ along the shortest path to node $j$ retains the greatest signal fidelity, we define the distortion function of any signal $x$ from brain region $i$ to a compressed representation $\hat{x}$ decoded in brain region $j$ as: 
	
	\begin{equation}
	d(x, \hat{x})_{i j}=(1-\eta),
	\end{equation}
	
	\noindent where $\eta$ denotes the probability that a walker gets from node $i$ to node $j$ along the shortest path. A signal with greater probability $\eta$ of propagating by the shortest path between brain region $i$ and brain region $j$ is at a lower risk of distortion (see \textbf{Figure \ref{fig4}D}). Intuitively, increased topological distance adds greater risk of signal distortion due to further transmission through capacity-limited channels (i.e. structural connections), temporal delay, and potential mixing with other signals. Given the measure of resources in Equation \ref{eq9}, we develop and test predictions of a novel definition of the rate $R(D)$; here, we define $R(D)$ as the resources $r_{i j}(\eta)$ required to achieve a tolerated level of distortion $d(x, \hat{x})_{i j}$:
	
	\begin{equation}
	R(D) \equiv r_{i j}(d(x, \hat{x})_{i j}),
	\end{equation}	 
	
	\noindent as in (\textbf{Figure \ref{fig4}D}). When the log of resources $\mathrm{log}(r_{i j})$ is plotted against our metric of distortion $D = d \in 1-\eta_{i j}$, the exponential gradient is depicted linearly (see \textbf{Figure \ref{fig4}E}). Because prior work focused on 50\% distortion during analyses, we required the slope to intersect the mean midpoint rate at 50\% distortion \cite{goni2013exploring}. In addition to the precedent offered by prior work, this requirement is also reasonable given that we sought to model both high and low distortions equitably. The slope denotes the minimum number of resources required to achieve a tolerated level of distortion, which we refer to as the \emph{compression efficiency} (\ref{fig4}E; bottom). A steeper slope (i.e., a more negative relation) reflects reduced compression efficiency, or prioritization of message fidelity. A flatter slope (i.e., a more positive relation) reflects increased compression efficiency, or prioritization of lossy compression. Individual variation in compression efficiency can be assessed by using the average resource efficiency across brain regions. When compression efficiency is computed for sets of brain regions by averaging across individuals, the slope can denote either messages sent from or arriving to a brain region by using the average resource efficiency over either all nodes $j$ or all nodes $i$, respectively. \\
	
	\rev{While prior work calculates the bit rates that arise from the stochasticity of opening and closing ion channels or releasing a synaptic vesicle \cite{laughlin2001energy}, in this manuscript we specifically chose to focus on macroscale level neuroimaging data. Our choice stemmed in part from the fact that the surrounding background literature in neuroimaging motivated many components of our theory, and in part from the fact that the theory of efficient coding had not yet been extended to this scale. We sought to understand the role of the whole connectome in supporting efficient communication and information processing.} \\
		
    \rev{To calculate numerical measures of bit rates at any spatial scale, we require a probabilistic information source. In the random walk model of inter-regional communication, each brain region is an information source that activates while sending a message---that is, a random walker---to another region along the wiring of the connectome. The information content of an individual message is determined by the probability of a message being sent, which, in our setting, is equivalent to the probability of a brain region changing its activity (activating or deactivating) in statistical association with behaviors and cognitive functions. At the level of macroscale neuroimaging data, individual studies describe localized neural activation and deactivation associated with specific cognitive tasks and behavior. Meta-analyses of these individual studies can help aggregate and summarize a large quantity of data on the general probability of brain region activation across a wide range of behaviors and cognitive functions \cite{10.2307/j.ctt17kk982, Yarkoni2011LargescaleAS}. We used the meta-analytic approach of NeuroSynth to obtain probability distributions across statistical maps of neural activity \cite{Yarkoni2011LargescaleAS}. We then used this average probability that a brain region is active to calculate the information content of each communication event, or the transmission of one neural message.}\\
		
	\rev{To calculate the information content of a neural message, we used the Shannon information measure to calculate information rate in bits per second \cite{Stone2016PrinciplesON}. Shannon information derives from the probability of a particular random event. Intuitively, less probable events convey messages with more information content, and \textit{vice versa}. In the context of macroscale brain networks, we sought to obtain the probability of a brain region transmitting a neural signal measured by fMRI BOLD activation (see \textbf{Supplementary Figure 3}). The mean probability of regional activation then is used to calculate the information content per message as the Shannon information given by the following expression:}
	
	\begin{equation}\label{shannonInfo}
	\rev{H = -log_2(P(Activation|Terms)),}
	\end{equation}
	
	\rev{\noindent where $H$ denotes the information content, or surprisal, of a neural signal measured by fMRI, and $P(Activation|$ $Terms)$ defines the average conditional probability of brain region activation given all $>$3,200 psychological terms encompassing an ontology that spans sensation, behavior, cognition, emotion, and disorders \cite{Yarkoni2011LargescaleAS}. The average information content of a message, which we model with a random walker, is 5.74 bits. A characteristic timescale of fMRI activity is given by the repetition time (TR) of each measurement, which is 0.72 seconds in the Human Connectome Project neuroimaging sequence \cite{barch2013function} and can vary across studies. Hence, the information rate is 5.74 bits/0.72 seconds = 7.97 bits/second per channel use. As with the estimation of information content, the main results of our paper do not depend on the choice of a characteristic timescale, because all values are converted using the same unit of bits per message. In selected early graphs, we include a secondary y-axis to indicate the conversion between units of random walkers and units of bits. To obtain the average bits per transmission for different levels of distortion, we multiplied the number of messages (random walkers) by 5.74 bits/message. We note that our theory can be applied to different neural activity measurements with connectomes reconstructed from a range of spatial scales. In future work, it will be important to analyze the predictability of bit rates measured directly from neural recordings based on theoretical estimates derived from the connectome.}\\
	
	\subsubsection{Biased Random Walk} \label{biasedRW}
	
	Given the advantages of \rev{random walkers which propagate along the} shortest path, we sought to assess how brain metabolism could support the reliability and fidelity of signaling. \rev{Metabolic} diffusion can be modeled as random walks over a structural connectivity matrix biased by regional CBF \cite{alt1980biased}. Electrical signal propagation could be modeled as random walks over a structural connectivity matrix biased by regional intracortical myelin \cite{glasser2011mapping}. To model \rev{metabolic} diffusion of random walkers attracted to brain regions of high CBF, we used analytical solutions to biased random walks. First, we defined the matrix $\mathbf{T}$ of CBF-biased transition probabilities as:
	\begin{equation}\label{eq16}
	T_{i j}^{\alpha}=\frac{\alpha_{i j} A_{i j}}{\sum_{k} \alpha_{i k} A_{i k}},
	\end{equation}
	\noindent where the element of $T_{i j}$ defines the transition probabilities of a random walker traversing edges of the structural connectivity matrix $\mathbf{A}$ which are multiplied by a bias term $\alpha$. For random walkers attracted to brain regions of high CBF, the bias term $\alpha$ was defined as the average CBF value for each pair of brain regions. Hence, a random walker propagates over the brain's structural connections with transition probabilities of $T_{i j}$ that reflect the integrity of structural connections and the average level of CBF between pairs of brain regions. We then substituted the $\mathbf{U}$ matrix in the resources $r_{i j}(\eta)$ of Equation \ref{eq9} with $T_{i j}$ in Equation \ref{eq16} to compute the number of resources required for a biased random walker to propagate by the shortest path with a specified probability. \rev{To model electrical signal propagation, we performed the same matrix transformation as described above, using regional measurements of intracortical myelin.}
	
	\subsubsection{Rich Club}
	
	Due to the importance of brain network hubs in the broadcasting of a signal \cite{bullmore2012economy,van2013network,avena2018communication}, we sought to identify the set of high-degree brain regions in the rich club (see \textbf{Figure \ref{fig8}A}) \cite{colizza2006detecting}. To identify the subnetwork of rich club brain regions, we computed the weighted rich club coefficient $\Phi^{z}(k)$ as:
	\begin{equation}\label{eq17}
	\phi^{z}(k)=\frac{Z_{>k}}{\sum_{l=1}^{E>k} z_{l}^{\text { ranked }}},
	\end{equation}
	\noindent where $Z^{ranked}$ is a vector of ranked network weights, $k$ is the degree, $Z_{>k}$ is the set of edges connecting the group of nodes with degree greater than $k$, and $E_{>k}$ is the number of edges connecting the group of nodes with degree greater than $k$. Hence, the rich-club coefficient $\Phi^{z}(k)$ is the ratio between the set of edge weights connected to nodes with degree greater than $k$ and the strongest $E_{>k}$ connections. The rich-club coefficient was normalized by comparison to the rich club coefficient of random networks \cite{colizza2006detecting}. Random networks were created by rewiring the edges of each individual's brain network while preserving the degree distribution. The rich-club coefficient for the randomized networks $\Phi_{\text { random }}(k)$ was computed using Equation \ref{eq17}. Then, the normalized rich-club coefficient $\Phi_{\text { norm }}(k)$ was calculated as follows:
	\begin{equation}
	\phi_{\mathrm{norm}}(k)=\frac{\phi(k)}{\phi_{\mathrm{random}}(k)},
	\end{equation}
	\noindent where $\Phi_{\text { norm }}(k)>1$ indicates the presence of a rich club organization. We tested the statistical significance of $\Phi_{\text { norm }}(k)$ using a 1-sample $t$-test at each level of $k$, with family-wise error correction for multiple tests over $k$. Each individual was assigned the value of their highest degree $>k$ rich club level and their nodes were ranked by rich club level. Over the group of individuals, the nodal ranks were averaged and the top 12\% of nodes were selected as the rich club, following prior work \cite{collin2013structural}.
	
	\subsubsection{\rev{Hierarchical Organization}}
	
	\rev{Hierarchical modularity is a form of network complexity that is characterized by small groups of nodes organized successively into increasingly larger groups. Networks with hierarchical organization exhibit a distinctive scaling where transitivity (clustering coefficient) decreases exponentially as a function of degree \cite{ravasz2003hierarchical}. This scaling appears linear in a log-log plot and the slope will be used to assess the hierarchical organization as the hierarchy parameter $\zeta$. To assess hierarchical structure, we examined the relationship between the nodal transitivity (clustering coefficient) and nodal degree \cite{ravasz2003hierarchical}:}
	\begin{equation}
	\rev{C(k)=k^\zeta},
	\end{equation}
	\noindent \rev{where $C$ is the transitivity (clustering coefficient), $k$ is the degree, and $\zeta$ represents the extent of hierarchical organization. To estimate $\zeta$, we used similar curve-fitting methods as in estimating the compression efficiency gradient. The transitivity $C$ was defined as:}
	\begin{equation}
	\rev{C_i=\frac{2n_i}{k_i(k_i-1)}},
	\end{equation}
	\rev{where $n$ is the number of links between the $k$ neighbors of node $i$. Intuitively, a high $\zeta$ value indicates hierarchical organization with less clustering about central nodes (with higher degree), and a low $\zeta$ value indicates hierarchical organization with more clustering about central nodes (with higher degree).}
	
	\subsection{Network Null Models} \label{networkNulls}
	
	Random graphs are commonly used in network science to test the statistical significance of the role of some network topology against null models. We used randomly rewired graphs generated by shuffling each individual's empirical networks 20 times, as in prior work \cite{maslov2002specificity}. Furthermore, we generated Erd\H{o}s-Reny\'i random networks for each individual brain network where the presence or absence of an edge was generated by a uniform probability calculated as the density of edges existing in the corresponding brain network. Edge weights were randomly sampled from the edge weight distribution of the brain network. While the randomly rewired graphs retain empirical properties such as the degree and edge weight distributions of the individual brain networks, the Erd\H{o}s-Reny\'i networks do not. Hence, the randomly rewired null network was used in all analyses where the degree distribution should be retained (e.g., normalized rich club coefficient), while the Erd\H{o}s-Reny\'i network was used in analyses assessing the overall contribution of the brain network topology (e.g., compression efficiency). ~\\ 
	
	\revtwo{We used the Erd\H{o}s-Reny\'{i} networks as the null network for two reasons. First, Erd\H{o}s-Reny\'{i} networks are particularly appropriate for comparing the random walk model to the alternative shortest path routing model. Prior work found that Erd\H{o}s-Reny\'{i} were networks with very high global efficiency, though the networks were biologically implausible due to their connection costs \cite{sporns2013network}. Thus, the Erd\H{o}s-Reny\'{i} networks serve as a benchmark for a communication architecture supporting highly efficient shortest path routing.}\\
		
	\revtwo{Second, we chose to use Erd\H{o}s-Reny\'{i} networks because we sought to compare random walk communication on brain network connectivity to random walk communication on synthetic networks that optimally minimize our distortion function---that is, maximizing the shortest path probability \cite{goni2013exploring}. In doing so, we can address the major criticism of the plausibility of random walk models: inefficiency \cite{avena2018communication}. Prior work compared random walks dynamics on canonical graph models, finding that the Erd\H{o}s-Reny\'{i} networks had the highest probability of shortest path propagation \cite{goni2013exploring}. Thus, we expected that the Erd\H{o}s-Reny\'{i} networks would approximate a limit on the efficiency for random walk communication dynamics according to the rate-distortion model. Compared to this approximate limit of efficiency, we aimed to show that biological investments in brain network communication modeled using random walks could indeed be efficient. Moreover, we aimed to show that inefficiency, when viewed in the light of the biologically-established efficient coding framework, uses the redundancy of communication to overcome noise introduced by intrinsic stochasticity of neural processes to balance the transmission fidelity and lossy compression of information \cite{barlow1961possible}.}\\
	
	Our tests using the randomly rewired network evaluate the null hypothesis that an apparent rich-club property of brain networks is a trivial result of topology characteristic of random networks with some empirical properties preserved, as in prior work \cite{colizza2006detecting,van2013network}. The alternative hypothesis is that the brain network has a rich-club organization beyond the level expected in the random networks. Our tests using the Erd\H{o}s-Reny\'i network evaluate the null hypothesis that the rates in the rate-distortion function modeling information processing capacity in brain networks does not differ from the rates in the rate-distortion function of random networks. The alternative hypothesis is that the rate of the brain network's rate-distortion function differs from that of random networks, consistent with the notion that Erd\H{o}s-Reny\'i networks have a greater prevalence of shortest paths compared to brain networks. We additionally used the Erd\H{o}s-Reny\'i network to assess the hypothesis of rate-distortion theory that synthetic networks should exhibit the same information processing trade-offs (the monotonic rate-distortion gradient) as empirical brain networks \cite{sims2018efficient}. We selected Erd\H{o}s-Reny\'i networks to assess these hypotheses for two reasons. First, Erd\H{o}s-Reny\'i networks do not retain core architectures of brain networks, such as modularity, and therefore reflect an extreme synthetic network. Second, Erd\H{o}s-Reny\'i networks are commonly used as a benchmark for assessing shortest path prevalence due to the prominence of uniformly distributed direct pairwise connections \cite{sporns2013network, avena2018communication}. In light of the central assumption that shortest paths represent the route of highest signal fidelity in our definition of distortion, we used Erd\H{o}s-Reny\'i networks to verify our intuition that compression efficiency should be greater in the Erd\H{o}s-Reny\'i network than in brain networks. \\
	
	\rev{Scale-free networks have a degree distribution resulting in a subset of highly connected hubs. Prior work established that global transitivity decreases exponentially with network size in a scale-free network, but that networks that are scale-free \textit{and} have hierarchical organization can decouple transitivity from size. We created non-hierarchical scale-free networks using the Barab\'asi-Albert algorithm (with the preferential attachment parameter set to 1 for linear preferential attachment) and re-evaluated whether the structural brain networks with hierarchical organization similarly decoupled transitivity and size \cite{barabasi1999emergence}. Decoupling transitivity and size allows for simultaneously high levels of transitivity despite large size. In the context of efficient coding theory, we chose to compare the global transitivity of structural brain networks with scale-free networks across 5 network sizes to examine if hierarchical organization in the human structural brain networks confers a similar outcome.}

	\subsection{Statistical Analyses}
	
	To assess the covariation of our measurements across individuals and brain regions, we used generalized additive models (GAMs) with penalized splines. GAMs allow for statistically rigorous modeling of linear and non-linear effects while minimizing over-fitting \cite{wood2004stable}. Throughout, the potential for confounding effects was addressed in our model by including covariates for age, sex, age-by-sex interaction, network degree, network density, and in-scanner motion. \rev{Due to the likelihood of inflated estimates of brain-behavior associations despite well-powered analyses \cite{Marek2020.08.21.257758}, we report bootstrap 95\% confidence intervals for the test statistic and adjusted $R^2$ of each model across 1000 bootstrap samples.} 
	
	\subsubsection{Metabolic running costs associated with brain network architectures}
	
	We used penalized splines to estimate the nonlinear developmental patterns of global efficiency (Equation \ref{globEffEq}) and CBF, as in prior work \cite{satterthwaite2014impact,baum2017modular}. Then, we assessed the partial correlation between the residual variance (unexplained by covariates of age, sex, age-by-sex-interaction, degree, density, and motion) of global efficiency and CBF. The final models can be written as:
	\begin{equation}
	\textrm{Global efficiency $\sim$ spline(age) + sex + spline(age by sex) + degree + density + motion},
	\end{equation}
	\begin{equation}
	\textrm{CBF $\sim$ spline(age) + sex + spline(age by sex) + degree + density + motion},
	\end{equation}
	\noindent and
	\begin{equation}
	\textrm{Residual(Global efficiency) $\sim$ residual(CBF)}.
	\end{equation}
	To evaluate the importance of age as a confound for the relationship between global efficiency and CBF, we also performed sensitivity analyses by removing selected covariates and re-assessing the model. In addition, for consistency with prior work \cite{varkuti2011quantifying}, we performed the same analysis including covariates for gray matter volume and density. ~\\
	
	To assess the relationship between CBF and the strength of structural paths supporting \rev{random walks} (Equation \ref{eqPathStrength}), we again used penalized splines. The final model can be written as:
	\begin{equation}
	\textrm{CBF $\sim$ path strength + spline(age) + sex + spline(age by sex) + degree + density + motion}.
	\end{equation}
	Assessments of path strengths were corrected for false discovery rate across the statistical tests performed over the discrete path lengths. 
	
	\subsubsection{Trade-offs between modularity and \rev{random walk} architecture}
	
	Next, we sought to evaluate the metabolic running cost of brain network properties, in line with calls for investigation of the economic landscape of resource-constrained trade-offs between hallmark brain network architectures such as modularity (Equation \ref{eqModularity}) and new measures of brain network organization \cite{bullmore2012economy}. Following our findings that CBF is associated with structural properties supporting \rev{random walks}, we investigated path transitivity (Equation \ref{eqPathTrans}). We continued to use penalized splines to model the non-linear patterns of CBF and brain properties of interest. The final model can be written as:
	\begin{equation}
	\begin{split}
	\textrm{CBF $\sim$ path transitivity + modularity + path transitivity by modularity}\\ 
		\textrm{+ spline(age) + sex + spline(age by sex) + degree + density + motion}.
	\end{split}
	\end{equation}
	To visualize the landscape of CBF as a function of modularity and path transitivity, we plotted the GAM model response function. We described the distribution of modularity and path transitivity across individuals using frequency histograms. \rev{We calculated a map of change in global CBF with respect to both modularity and path transitivity using first-order derivatives. A saddle point suggests that adaptive compromises in network architecture are constrained by dual objectives. To quantify the location of the saddlepoint coordinate within the change map, we performed a $k$-nearest neighbor search of the value 0 in the gradient of first-order derivatives indicating minima and maxima.}
	
	\subsubsection{Compression efficiency and development}
	
	To assess the possibility of distinct compression efficiency of brain networks compared to random networks, we calculated the resource efficiency (Equation \ref{eq9}) at 14 levels of distortion and performed an analysis of variance (ANOVA) test. The ANOVA model can be written as:
	\begin{equation}\label{eq24}
	\textrm{Resources $\sim$ distortion + type of network},
	\end{equation}
	where the type of the network is a categorical variable designating if the network was a brain network or a random network. ~\\
	
	To compute compression efficiency per individual brain network, we used a polynomial regression function to find the best linear fit to the monotonic rate-distortion function according to the prediction of a linear rate-distortion gradient in semi-log space (log(resources) as a function of distortion). Next, we used a GAM model to assess the non-linear patterns of compression efficiency in development, which we can formally write as follows:
	\begin{equation}
	\textrm{Compression efficiency $\sim$ spline(age) + sex + spline(age by sex) + degree + density + motion}.
	\end{equation}
	
	\rev{To quantify the differences between the resources calculated from resource efficiency and the compression efficiency across 14 levels of distortion, we performed two-sample $t$-tests while controlling for family-wise error rate across multiple comparisons. For each level of distortion, we calculated the $t$-statistic comparing all individual resources to all individual resources predicted by the linear rate-distortion gradient.} 
	
	\subsubsection{Compression efficiency of \rev{biased random walks}}
	
	To compute the compression efficiency of \rev{metabolic} diffusion \rev{and electrical signaling}, we modified the model of Equation \ref{eq24} to instead calculate resource efficiency using the biased random walk matrices from Equation \ref{eq16}. The model can be written:
	\begin{equation}
	\textrm{Resources $\sim$ distortion + type of random walk + distortion by type of random walk},
	\end{equation}
	\noindent where the type of random walk is a categorical variable designating unbiased random walks using the structural network, biased random walks using the structural network biased with CBF, and biased random walks using the structural network biased with \rev{intracortical myelin}. To assess the hypothesis that resources differ according to the type of random walk, we performed $t$-tests while controlling for family-wise error rate across multiple comparisons. ~\\
	
	\subsubsection{Compression efficiency in a low or high fidelity regime}
	
	Next, we sought to test the predictions of a high or low fidelity communication regime. In a high fidelity regime, minimum resources given an expected distortion should increase monotonically as a function of network complexity. To assess whether the relationship between resources and network complexity (operationalized here as network size) is monotonic, we used a linear model written as:
	\begin{equation}
	\textrm{Log$_2$(resources) $\sim$ log$_1$$_0$(network size)}.
	\end{equation}
	\noindent In a low fidelity regime, minimum resources given an expected distortion should plateau as a function of complexity. We hypothesized that path transitivity is a property of structural networks that supports lossy compression and storage savings. The complexity of the shortest path was defined as the number of nodes contributing to path transitivity (Equation \ref{eqPathTrans}). To assess whether the resources non-linearly plateau as a function of shortest path complexity, we used a GAM model written as follows:
	\begin{equation}
	\textrm{Log$_2$(resources) $\sim$ spline(shortest path complexity)}.
	\end{equation}
	
	\rev{In light of our interest in network size and shortest path complexity (path transitivity) as measurements of high and low fidelity regimes, we next sought to measure hierarchical organization. Networks that are scale-free and have hierarchical organization can achieve simultaneously high network size and transitivity \cite{ravasz2003hierarchical}. To assess whether hierarchical modularity supports a high fidelity regime, we used a linear model written as follows:}
	\begin{equation}
	\rev{\textrm{Log$_2$(resources) $\sim$ hierarchical modularity}}.
	\end{equation}
	
	\rev{We next evaluated whether brain structural networks with hierarchical modularity \textit{and} the scale-free property achieved greater transitivity than random networks with only the scale-free property, as predicted by prior findings \cite{ravasz2003hierarchical}. To evaluate the differences in transitivity according to the network properties of hierarchical organization and the scale-free property, we used the ANOVA model written as:}
	\begin{equation}
	\rev{\textrm{Global transitivity $\sim$ network size + type of network}},
	\end{equation}
	\noindent \rev{where the type of network is a categorical variable designating either the human brain structural network or the Barab\'asi-Albert scale-free network.}
	
	\subsubsection{Compression efficiency and patterns of neurodevelopment}\label{spatialPermutation}
	
	To explore how compression efficiency might relate to patterns of cortical myelination and areal scaling, we assessed the Spearman's correlation coefficient between myelination or scaling and send or receive compression efficiency. \rev{We also assessed the relationship between cortical areal scaling and nodal transitivity, in light of our hypothesis that greater path transitivity may support greater fidelity and the efficient coding hypothesis which states that the organization of the brain allocates neural resources according to the physical distribution of information.} To further test correspondence between brain maps, we used a spatial permutation test, which generates a null distribution of randomly rotated brain maps that preserve the spatial covariance structure of the original data \cite{alexander2018testing}. \revtwo{Using spatially-constrained null models is the state-of-the-art for comparing brain maps \cite{alexander2018testing}. We used a spin test variant that reassigns parcels with no duplication. This method implements an iterative procedure that uniquely assign parcels based on Euclidean distance of the rotated parcels, ignoring the medial wall and its location \cite{vavsa2018adolescent}. Our choice of null model introduces a trade-off between permitting slightly more liberal critical thresholds than other spatially-constrained null models and retaining the exact distribution of the original brain map to better test network-based statistics like compression efficiency and transitivity \cite{markello2021comparing}.} We refer to the $p$-value of this statistical test as $p_{SPIN}$. Finally, we applied the conservative Holm-Bonferroni correction for family-wise error across these tests. 
	
	\subsubsection{Compression efficiency and rich-club hubs}
	
	Given the assumed integrative and broadcasting function of rich-club hubs, we sought to evaluate whether compression efficiency differed in rich-club hubs compared to other brain regions. We used the Wilcoxon rank-sum test to compare regional compression efficiency of either receiving or sending messages. Moreover, we assessed whether there was a difference between CBF in the rich-club hubs compared to other brain regions. Lastly, we tested the correlation of compression efficiency in the rich-club hubs and other brain regions with cognitive efficiency. To model non-linear patterns of cognitive efficiency, we used penalized splines controlling for potentially confounding covariates. The final model can be written as:
	\begin{equation}
	\begin{split}
	\textrm{Cognitive efficiency $\sim$ compression efficiency + spline(age)} \\
		\textrm{+ sex + spline(age by sex) + degree + density + motion}.
	\end{split}
	\end{equation}
	Due to previous report of the relationship between cognition and global efficiency (Equation 1), we determined that compression efficiency and global efficiency were not collinear and therefore conducted a sensitivity analysis including global efficiency as a covariate. The model was written as:
	\begin{equation}
	\begin{split}
	\textrm{Cognitive efficiency $\sim$ compression efficiency + global efficiency} \\
		\textrm{+ spline(age) + sex + spline(age by sex) + degree + density + motion}.
	\end{split}
	\end{equation}
	
	\subsection{Code availability}
	
	Code can be found at \url{https://github.com/dalejn/economicsConnectomics}.
	
	\subsection{Data availability}
	
	Neuroimaging and cognitive test data were acquired from the Philadelphia Neurodevelopmental Cohort. The data reported in this paper have been deposited in the database of Genotypes and Phenotypes under accession number dbGaP: phs000607.v2.p2 (\url{https://www.ncbi.nlm.nih.gov/projects/gap/cgi-bin/study.cgi?study_id=phs000607.v2.p2}). \rev{The allometric cortical scaling maps were downloaded from  \href{https://neurovault.org/collections/3901/}{a NeuroVault repository} \cite{reardon2018normative} and \href{http://brainvis.wustl.edu/wiki/index.php/Caret:MyelinMaps}{the cortical myelin maps were downloaded from a public resource} \cite{glasser2011mapping}.} 
	
	\subsection{Citation Diversity Statement}
	
	Recent work in several fields of science has identified a bias in citation practices such that papers from women and other minority scholars are under-cited relative to the number of such papers in the field \cite{mitchell2013gendered,dion2018gendered,caplar2017quantitative, maliniak2013gender, Dworkin2020.01.03.894378, bertolero2021racial, wang2021gendered, chatterjee2021gender, fulvio2021imbalance}. Here we sought to proactively consider choosing references that reflect thediversity of the field in thought, form of contribution, gender, race, ethnicity, and other factors. First, we obtained the predicted gender of the first and last author of each reference by using databases that store the probability of a first name being carried by a woman \cite{Dworkin2020.01.03.894378,zhou_dale_2020_3672110}. By this measure (and excluding self-citations to the first and last authors of our current paper), our references contain 18.07\% woman(first)/woman(last), 5.75\% man/woman, 23.33\% woman/man, and 52.85\% man/man. This method is limited in that a) names, pronouns, and social media profiles used to construct the databases may not, in every case, be indicative of gender identity and b) it cannot account for intersex, non-binary, or transgender people. Second, we obtained predicted racial/ethnic category of the first and last author of each reference by databases that store the probability of a first and last name being carried by an author of color \cite{ambekar2009name, sood2018predicting}. By this measure (and excluding self-citations), our references contain 7.38\% author of color (first)/author of color(last), 12.18\% white author/author of color, 24.82\% author of color/white author, and 55.62\% white author/white author. This method is limited in that a) names and Florida Voter Data to make the predictions may not be indicative of racial/ethnic identity, and b) it cannot account for Indigenous and mixed-race authors, or those who may face differential biases due to the ambiguous racialization or ethnicization of their names.  We look forward to future work that could help us to better understand how to support equitable practices in science.
	
	\singlespacing
	
	\newpage
	
	\bibliographystyle{ieeetr}
	\bibliography{./bibfile}

\end{document}